# Nitrogen Vacancy Centers in Diamond for Quantum Biosensing: Magnetometry Techniques, Platforms and Applications

**Sirsendu Ghosal[1], Umakant Prajapati[1], Sachin Negi[1], Saifian Farooq Bhat[1], Thejas B[1] and Vibhav Bharadwaj[1*]**

[1]*Department of Physics, Indian Institute of Technology Guwahati, Guwahati 781039, India*

## Abstract

Quantum sensing using nitrogen-vacancy (NV) centers in diamond has emerged as a powerful platform for detecting ultra-low magnetic fields under ambient conditions. Owing to their long spin coherence times, optical addressability, and compatibility with aqueous environments, it has found widespread applications in biosensing and bio-imaging. This review presents the fundamental principles and recent advances in NV-based quantum magnetometry for biosensing applications, with a particular focus on measurements in aqueous medium and at cellular and molecular length scales. We discuss the underlying spin physics of NV centers and highlight two primary detection modalities: optically detected magnetic resonance (ODMR) and $T_1$ relaxometry-based sensing and how these approaches aids in the detection of both static magnetic fields and dynamic magnetic noise arising from biological processes. The review explores key application areas, including nanoscale nuclear magnetic resonance (NMR), monitoring of neural activity, detection of abnormal or rogue cells using NV-based platforms etc. In addition, strategies for enhancing sensitivity, such as surface functionalization of nanodiamonds, femtosecond (fs) laser-written photonic structures, and integration with microfluidic and lab-on-chip systems have also been discussed in depth. We also address the critical challenges, including surface-induced decoherence, charge-state instability, and signal-to-noise limitations in biofluids associated with NV-based biosensing applications. Finally, we outline future prospects, highlighting how NV-based magnetic biosensing provides a promising pathway for translating quantum sensing technologies into practical biomedical applications.

[*] Corresponding author; email: vibhav9@iitg.ac.in

# 1 Introduction

The convergence of quantum science and biomedical research has created unprecedented opportunities for developing sensing technologies capable of probing biological systems with exceptional sensitivity and spatial resolution.[1,2,3,4] Negatively charged nitrogen vacancy centers ($NV^-$) in diamond has emerged as one of the versatile platforms among all other solid-state quantum sensors for magnetic biosensing owing to its outstanding quantum properties, including long spin coherence times, room-temperature functionality, optical initialization and readout capabilities, and inherently high biocompatibility. Unlike other conventional magnetic sensing technologies such as superconducting quantum interference devices (SQUIDs), Optically pumped magnetometry (OPM) or Magnetic Resonance Imaging (MRI) etc.[5] which often requires heavy instrumentation or cryogenic conditions to operate, NV-center-based magnetic sensors can operate in ambient conditions and can perform non-invasive measurements in aqueous and physiological environments.[6,7] Due to these unique attributes, diamond quantum sensors are especially well-suited for studying biological processes at molecular, cellular, and tissue scales.[8]

The NV center is formed by a substitutional nitrogen atom sitting next to a vacancy in the diamond lattice. Its electron spin can be optically initialized and read out via spin-dependent fluorescence, while microwaves enable coherent control of the spin state.[9] This combination forms the basis of optically detected magnetic resonance (ODMR), allowing highly sensitive measurements of local magnetic fields, temperature, electric fields, strain, and pressure.[10] A second sensing approach, spin relaxometry, relies on the spin relaxation dynamics of NV centers and is especially responsive to fluctuating magnetic fields from paramagnetic species like free radicals and transition-metal` ions etc.[11] Together, ODMR and relaxometry provide complementary ways to detect both static fields and dynamic magnetic noise linked to biological activity.

Recent advances in diamond growth, nanofabrication, and surface engineering have significantly expanded the scope of NV based biosensing applications.[12,13] Single-crystal diamond chips with shallow NV ensembles serve as sensitive platforms for wide-field magnetic imaging of biological samples, while fluorescent nanodiamonds (FNDs) - owing to their small size, strong photostability, and low toxicity to cells - allow quantum sensing to be carried out inside living cells.[14,15,16] Additionally, functionalization of diamond surfaces with antibodies, peptides, aptamers, or nucleic acids adds the ability to target specific cells, organelles, or biomolecules, paving the pathway for applications like molecular diagnostics, targeted drug delivery, and intracellular monitoring etc.[17]

Due to the rapid technological development, it has led to a rapidly growing range of biomedical applications. From monitoring neuronal activity to targeted drug delivery to imaging of biomolecules, NV center based quantum magnetometry is rapidly expanding its horizons in the field of medical diagonistics.[17,18] More recently, hybrid diamond platforms- combining photonic structures, microfluidics, and standard semiconductor technologies have boosted photon collection efficiency, sensing throughput, and device portability - moving the field closer to practical, point-of-care quantum biosensors for clinical use.[19,20,21]

Despite its remarkable advances, several challenges still need to be addressed before NV-based quantum sensors can be widely translated into biomedical practice. Considerable research efforts are therefore directed toward improving diamond material quality, engineering robust surface chemistries, developing advanced quantum control protocols to overcome these limitations.[22,23]

In this review, we present a comprehensive overview of the NV-center-based magnetometry for biosensing. We begin with the fundamental spin physics of NV centers and the principles underlying ODMR and relaxometry, then survey the major classes of diamond quantum sensors and highlight their uses in relevant fields. Then we have briefly discussed about the state of the art major biosensing application areas where NV center based diamond magnetometry is making a major leap forward. We conclude by highlighting the inherent bottlenecks and emerging strategies for improving sensor performance, and by outlining future directions for translating diamond quantum sensing from laboratory research into real-world biomedical diagnostics and precision healthcare.

# 2 Principles of NV Center Based Quantum Sensing

Structurally, the NV defect arises from a nitrogen atom situated next to a vacant lattice site within the diamond host, as shown schematically in **Fig. 1(a)**. In its negatively charged state ($NV^{-}$), the defect exhibits $C_{3V}$ symmetry and can align along any of four equivalent ⟨111⟩ crystallographic directions. Because the spin splitting depends on the component of an external magnetic field along the NV axis, measuring this splitting for each orientation allows full vector reconstruction of the field - the basis of NV-based vector magnetometry. The ground-state spin triplet is governed by a spin Hamiltonian of the form[9]

$$\hat{H} = D\hat{S}_z^2 + \gamma_e \boldsymbol{B}.\hat{\boldsymbol{S}} + d_{\parallel} E_z \hat{S}_z^2 + d_{\perp}(E_x(\hat{S}_x^2 - \hat{S}_y^2) + E_y(\hat{S}_x\hat{S}_y + \hat{S}_y\hat{S}_x) + \hat{H}_{strain} \quad (1)$$

Here, $D$ = 2.87 GHz denotes the zero-field splitting parameter, $\gamma_e$ = 28 GHz.T$^{-1}$ is the electron gyromagnetic ratio, $\boldsymbol{B}$ is the applied external magnetic field, and $E$ represents the local electric field. The last term, $\hat{H}_{strain}$, captures perturbations arising from lattice strain. Since the Zeeman term typically dominates over the electric-field and strain contributions in most magnetometry applications, the Hamiltonian can be reduced to the simplified form.[8,24]

$$\hat{H} \approx D\hat{S}_z^2 + \gamma_e \boldsymbol{B}.\hat{\boldsymbol{S}} \quad (2)$$

The quantum sensing mechanism of the NV center is governed by its unique electronic energy-level structure, as illustrated in **Fig. 1(b)**. The $NV^-$ center possesses a spin-triplet ground state ($^3A_2$) comprising three spin sublevels, ($m_s$ = 0) and ($m_s$ = ±1). In the absence of an external magnetic field, the ($m_s$ = ±1) states are degenerate and are separated from the ($m_s$ = 0) state by the zero-field splitting, (D = 2.87) GHz. Upon optical excitation with a 532 nm green laser, electrons are promoted from the ground triplet ($^3A_2$) to the excited triplet ($^3E$). This optical excitation does not directly perform the sensing operation; rather, it prepares the NV center for spin-state initialization and optical readout.[25,26]

Following excitation, the electron can relax to the ground state through two distinct pathways. Electrons excited from the ($m_s$ = 0) state predominantly return directly to the corresponding ground state via a radiative transition, emitting red fluorescence centred around 637 nm. Owing to its higher fluorescence intensity, the ($m_s$ = 0) state is referred to as the bright state. In contrast, electrons occupying the ($m_s$ = ±*1*) states preferentially undergo a non-radiative intersystem crossing (ISC) through the metastable singlet states ($^1A_1$) and ($^1E$) before relaxing back to the ground state.[27,28] As this pathway suppresses fluorescence, the ($m_s$ = ±*1*) states are commonly referred to as the dark states. Importantly, the metastable singlet states relax preferentially into the ($m_s$ = *0*) ground state. Consequently, continuous optical excitation gradually pumps the electron population into the ($m_s$ = *0*) state, a process known as optical spin initialization or spin polarization. This optical pumping prepares the NV center in a well-defined quantum state, providing the essential starting point for quantum sensing measurements.[29]

Once the NV spins have been initialized, external perturbations such as magnetic fields modify the spin-state dynamics, which can be detected through changes in the optical signal. Based on the measured spin response, NV-center quantum sensing primarily employs two complementary detection modalities: ODMR, which measures shifts in the spin resonance frequencies induced by static or slowly varying magnetic fields, and spin relaxometry, which probes changes in the longitudinal spin relaxation time ($T_1$) arising from fluctuating magnetic fields and magnetic noise generated by dynamic biological processes**.** These two techniques form the foundation of most NV-center-based quantum sensing and biosensing applications.

## 2.1 Optically Detected Magnetic Resonance (ODMR)

To manipulate the spin state, a microwave (MW) field is applied via an integrated micro-antenna. When the MW frequency matches the zero-field $D_g = 2.87\ GHz$ or the Zeeman-split frequencies $(D_g \pm \gamma_{NV} B_0)$ it drives magnetic dipole transitions between the sublevels. This resonant field transfers population from $|m_s = 0\rangle$ state to the darker $|m_s = \pm 1\rangle$ states. By sweeping the MW frequency across these resonance conditions, a characteristic dip in the photoluminescence (PL) spectrum is produced as shown in **Fig 1(c)** providing an entirely optical readout of the magnetic resonance spectrum.[30] Depending on the measurement requirement, microwave manipulation is executed via two distinct modalities: Continuous-Wave (CW) ODMR: The green laser and the MW field are applied simultaneously. This is simpler to implement but suffers from power broadening of the resonance line due to constant laser repumping. Pulsed ODMR: The laser and MW fields are separated in time using pump-probe sequences. A laser pulse first initializes the spin, followed by precise MW pulses (like $\pi$ or $\frac{\pi}{2}$ rotations) during a dark period, and a final short laser pulse reads out the final state. This eliminates power broadening and allows high-sensitivity coherent control sequences like Hahn echo or dynamical decoupling.[10,31]

In the presence of an external magnetic field, the degeneracy of the $|m_s = \pm 1\rangle$ states is lifted by the Zeeman interaction, shifting the resonance frequencies according to

$$f_{\pm} = D_g \pm \gamma_e B_{\parallel} \tag{3}$$

$B_{\parallel}$ is the magnetic-field component projected along the NV symmetry axis. The corresponding resonance splitting is

$$\Delta f = f_{+} - f_{-} = 2\gamma_e B_{\parallel} \tag{4}$$

Thus, the external magnetic field can be directly determined by measuring the shift in the ODMR resonance frequencies. The magnetic sensitivity of an ODMR sensor is approximately given by

$$\eta_\beta = \frac{\Delta\nu}{\gamma_e C\sqrt{R}} \tag{5}$$

where $\Delta\nu$ is the ODMR linewidth, is the $C$ fluorescence contrast, and $R$ is the detected photon count rate. Therefore, narrower resonance linewidths, higher fluorescence contrast, and improved photon collection efficiency directly enhance the magnetic-field sensitivity.[32,33]

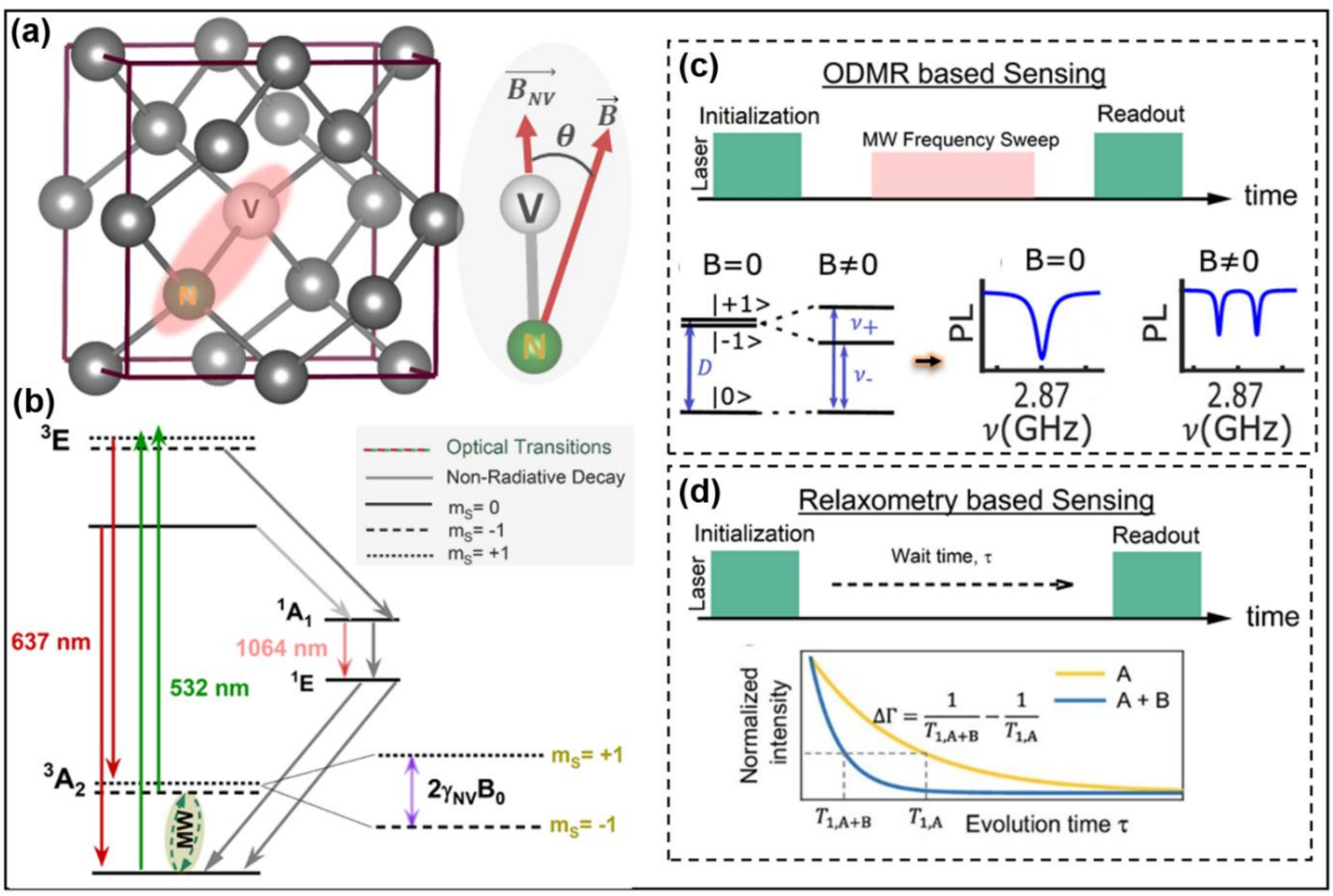

**Fig. 1.** (a) Atomic structure of an NV center in diamond, consisting of a substitutional nitrogen atom adjacent to a lattice vacancy. The right panel illustrates the orientation of the NV axis relative to the external magnetic field B, where θ represents the angle between the field and the NV symmetry axis. (b) Simplified energy-level diagram of the negatively charged NV center ($NV^-$), showing optical excitation at 532 nm, fluorescence emission at 637 nm, spin-dependent intersystem crossing through metastable singlet states, and Zeeman splitting of the $m_s = \pm1$ spin sublevels under an applied magnetic field. (c) Principle of ODMR-based sensing, where laser initialization and MW frequency sweeps lead to magnetic-field-dependent splitting of the resonance spectrum. (d) Relaxometry-based sensing scheme using longitudinal spin relaxation measurements, where environmental magnetic noise modifies the NV spin relaxation dynamics and fluorescence decay profile.

ODMR is particularly well-suited for detecting static (DC) and slowly varying (AC) magnetic fields, which has established it as the leading technique for vector magnetometry, wide-field magnetic imaging, current-density mapping, and temperature sensing. Within biological research, ODMR has found broad application in monitoring neuronal and cardiac action potentials, performing wide-field cellular magnetic imaging, tracking magnetically labelled cells and magnetic nanoparticles, and generating high-resolution magnetic maps of biological tissues.[34,35]

## 2.2 Spin Relaxometry

Despite the exceptional sensitivity of ODMR towards static and slowly varying magnetic fields, many biological processes generate rapidly fluctuating magnetic fields rather than coherent static signals. Such stochastic magnetic noise originates from unpaired electron spins associated with reactive oxygen species (ROS), free radicals, transition-metal ions, ferritin, and other paramagnetic biomolecules. To probe these dynamic magnetic environments, spin relaxometry has emerged as a powerful complementary sensing modality that monitors changes in the longitudinal spin relaxation time ($T_1$) of the NV center, instead of tracking resonance frequency shifts.[36,37]

The measurement protocol for spin relaxometry is illustrated in **Fig. 1(d)**.[11,38] Initially, the NV center is optically initialized into the $m_s=0$ state using a short pulse of 532 nm laser excitation. Following initialization, the laser is switched off for a variable dark interval, τ, during which the spin population relaxes towards thermal equilibrium through interactions with fluctuating magnetic fields in the surrounding environment. These magnetic fluctuations induce transitions between the $m_s = 0$ and $m_s = \pm 1$ spin states, thereby accelerating the spin relaxation process. Subsequently, a second laser pulse is applied to read out the remaining spin polarization via spin-dependent fluorescence. The measured fluorescence intensity follows an exponential decay given by

$$I(\tau) = I_0 \exp(-\frac{\tau}{T_1}) \qquad (6)$$

where $I(\tau)$ is the fluorescence intensity measured after the dark interval $\tau$, $I_0$ is the initial fluorescence intensity, and $T_1$ is the longitudinal spin relaxation time. The corresponding relaxation rate is expressed as

$$\Gamma_1 = \frac{1}{T_1} \tag{7}$$

which is directly proportional to the magnetic noise spectral density at the NV resonance frequency

$$\Gamma_1 \propto S_B(\omega_{NV}) \tag{8}$$

where $S_B(\omega_{NV})$ represents the spectral density of magnetic field fluctuations experienced by the NV center. Consequently, an increase in the local concentration of paramagnetic species enhances the magnetic noise, shortens the $T_1$ relaxation time, and produces a measurable reduction in the fluorescence signal. Owing to its high sensitivity to nanoscale magnetic fluctuations, spin relaxometry has become an indispensable technique for label-free detection of reactive oxygen species, molecular biomarkers, free radicals, intracellular redox processes, and transition-metal ions, enabling real-time monitoring of dynamic biochemical processes under physiological conditions.

# 3 Diamond Platforms for Bio-sensing

Depending on the intended application, NV centers can be incorporated into different diamond platforms ranging from millimeter-sized single-crystal substrates to nanoscale FNDs. Each platform offers distinct advantages in terms of magnetic sensitivity, spatial resolution, imaging depth, and compatibility with biological environments. Consequently, selecting an appropriate diamond platform has become a critical design consideration for next-generation quantum biosensors.

## 3.1 Single-Crystal Bulk Diamonds

Single-crystal electronic-grade diamond is widely regarded as the benchmark material for high-performance quantum sensing because it offers the longest NV spin coherence times, the lowest defect density, and exceptional optical quality.[39] In contrast to nanodiamonds, which often suffer from surface-induced spin decoherence, Brownian rotational motion, and random crystal orientations, bulk single-crystal diamond provides a stable crystallographic framework that enables highly reproducible measurements with prolonged spin coherence times ($T_2$).[40] High-purity electronic-grade diamonds are routinely synthesized by chemical vapor deposition (CVD) using ultra-low concentrations of paramagnetic impurities, resulting in millisecond-scale spin coherence times under dynamical decoupling sequences.[41,42] These extended coherence times translate directly into enhanced magnetic-field sensitivity, making bulk diamond the preferred platform for precision quantum magnetometry and wide-field magnetic imaging.

For biosensing applications, single-crystal diamond provides a robust and biocompatible sensing platform in which shallow NV ensembles are typically created 5-20 nm below the diamond surface through low-energy nitrogen ion implantation followed by high-temperature annealing.[43] This shallow placement ensures close proximity between the sensing layer and external biological samples while preserving excellent spin and optical properties.[44] Furthermore, the diamond surface can be chemically functionalized to support long-term growth and adhesion of living cells, including neurons and cardiomyocytes, without compromising cellular viability.[45,46] Such intimate contact enables efficient detection of weak magnetic fields generated by biological systems.

Bulk diamond also supports wide-field quantum imaging through dense ensembles of shallow NV centers that are simultaneously excited by a green laser and imaged using a scientific CMOS (sCMOS) camera.[47,48] This configuration enables high-speed, high-spatial-resolution imaging over a large field of view, allowing simultaneous visualization of entire neuronal networks, cardiac cell monolayers, magnetotactic bacteria, magnetic nanoparticles, and magnetically labelled cells under physiological conditions.[49,50]

Another significant advantage of bulk diamond is its compatibility with a broad range of quantum sensing modalities, including continuous-wave and pulsed ODMR, $T_1$ relaxometry, AC magnetometry, and quantum noise spectroscopy. Because the magnetic field generated by a source decays rapidly with distance (approximately as $1/r^3$), positioning the shallow NV sensing layer only a few nanometres below the surface minimizes signal attenuation from external biological samples.[51] This enables quantitative reconstruction of magnetic-field distributions and current-density maps with high spatial resolution. These capabilities have been extensively exploited for imaging neuronal activity, monitoring cardiac electrophysiology, and investigating magnetically active biological systems.[52] In addition, the large sensing area of bulk diamond allows simultaneous observation of thousands of cells, making it an exceptionally powerful platform for high-throughput bioimaging, electrophysiological studies, and real-time investigations of complex biological processes.[53,54]

For instance, Davis *et al*. developed a bulk single-crystal NV-diamond magneto-microscope to directly connect microscale magnetic field patterns in biological samples to macroscale MRI contrast. The platform used a CVD-grown, [100]-oriented bulk diamond chip with a relatively thick (~4 μm) near-surface NV layer, enabling wide-field, high-sensitivity vector magnetometry with reduced laser power and phototoxicity compared to shallow-implanted diamond sensors. The authors performed vector magnetometry on macrophages labelled with iron oxide nanoparticles (IONs) and on cryosectioned liver tissue from a mouse model of hepatic iron overload, mapping subcellular and sub-voxel magnetic dipole distributions with sub-micron spatial resolution and nT sensitivity. **(Fig. 2(a))**[55]

Despite these advantages, bulk diamond platforms are less suitable for intracellular sensing because the sensing element remains external to the biological sample. The magnetic field decreases rapidly with distance from its source, limiting sensitivity to deeply embedded intracellular processes.[40] Consequently, while bulk diamonds remain the preferred platform for tissue-scale magnetic imaging and functional bioelectromagnetic studies, nanoscale biological sensing increasingly relies on FND probes that can be delivered directly into living cells.

## 3.2 Fluorescent Nanodiamonds (FNDs)

FNDs have emerged as one of the most promising platforms for intracellular quantum biosensing by combining the exceptional spin properties of NV centers with the unique advantages of nanoscale particulate probes.[56,57,58,59] Unlike bulk diamond sensors, where the biological specimen remains external to the sensing element, FNDs can function as mobile intracellular quantum sensors owing to their nanoscale dimensions. These particles are chemically inert, highly biocompatible, and non-cytotoxic, offering a significant advantage over many conventional semiconductor quantum dots.[60] Living cells readily internalize FNDs through endocytic pathways, after which they remain stable without adversely affecting cellular viability, gene expression, or normal physiological functions. Consequently, FNDs are well suited for long-term intracellular sensing, in vivo tracking, and biomedical imaging applications.[61,62]

Typically ranging from 10 to 200 nm in diameter, FNDs retain many of the desirable properties of bulk diamond, including exceptional chemical stability, photostability, and resistance to photobleaching.[63] Their fluorescence remains stable over extended periods, enabling continuous tracking of individual particles inside living cells for days or even weeks.[64] In addition to their role as intracellular sensors, their excellent biocompatibility and ease of surface functionalization make them attractive candidates for targeted drug delivery and theranostic applications.[65,66,67]

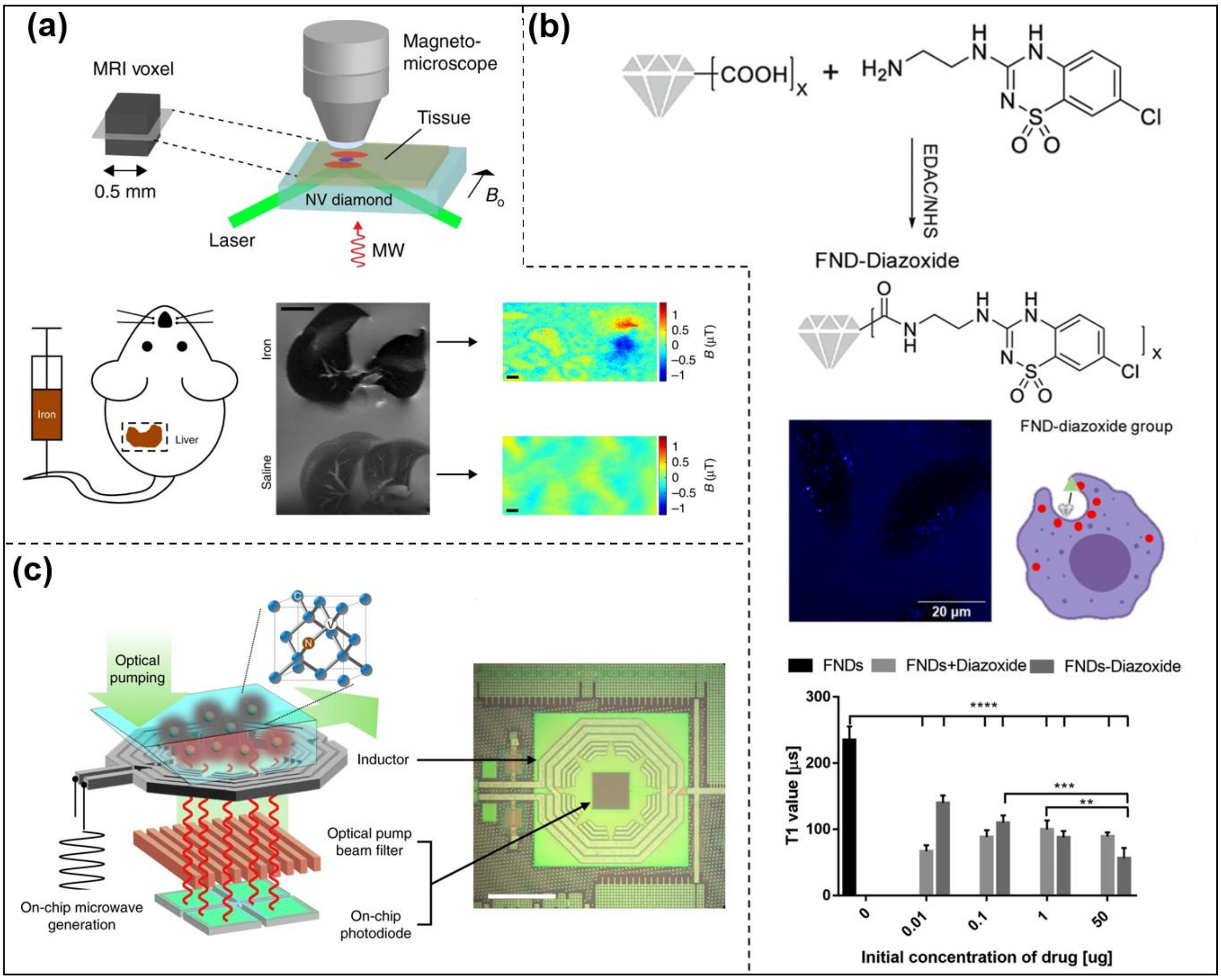

**Fig. 2.** Diamond platforms for biosensing and quantum bioimaging. (a) Single-crystal bulk diamond platform: Schematic of an NV-diamond magneto-microscope for sub-voxel magnetic field mapping and quantum imaging. *Adapted from Ref. [55]* (b) Fluorescent nanodiamond (FND) platform: Schematic illustration of the EDC/NHS-mediated conjugation of diazoxide to carboxyl-functionalized FNDs, forming FND-diazoxide conjugates. Confocal fluorescence imaging demonstrates intracellular localization of the functionalized FNDs in HeLa cells, while the schematic illustrates their cellular uptake and localized drug delivery. *Adapted from Ref. [75]* (c) CMOS-integrated NV sensor combining on-chip microwave, optical filtering, and photodetection beneath an NV-diamond slab, with side-coupled excitation enabled by a 45°-cut edge *Adapted from Ref. [83]*

One of the most important applications of FNDs is the quantum sensing of intracellular biochemical processes. Fluctuating magnetic fields generated by unpaired electron spins in reactive oxygen species (ROS) modify the longitudinal spin relaxation time ($T_1$) of nearby NV centers, facilitating real-time monitoring of oxidative stress, inflammation, mitochondrial dysfunction, and cellular ageing.[68] Furthermore, the spin resonance spectrum of NV centers is sensitive to local electric fields, pH-dependent surface charge variations, and membrane potential changes, allowing FNDs to probe dynamic intracellular environments with nanoscale spatial resolution. These capabilities have established FNDs as versatile quantum sensors for investigating complex biological processes that are difficult to access using conventional optical probes.[69]

The fabrication of FNDs generally begins with high-pressure high-temperature (HPHT) or chemical vapor deposition (CVD)-grown diamond nanoparticles.[70,71] More recently, cost effective bottom-up approaches like hydrothermal method has also been utilized to synthesize these FNDs.[72] NV centers are subsequently generated by high-energy electron irradiation or nitrogen ion implantation, which create vacancies within the diamond lattice.[73,74] High-temperature annealing, typically performed between 800°C and 1000°C, promotes vacancy diffusion and recombination with substitutional nitrogen atoms to form stable NV centers.[71] Post-processing steps, including oxidative and acid cleaning, remove graphitic surface layers produced during irradiation, improve colloidal stability, and introduce chemically active surface groups for subsequent bioconjugation. Recent advances in irradiation and annealing protocols introduced increasingly precise control over NV concentration, particle size, fluorescence brightness, and defect density, all of which critically influence the quantum sensing performance of FNDs.

To demonstrate the extraordinary capabilities of these FNDs, Tian *et al*. developed a multifunctional FND-based theranostic platform in which NV-containing fluorescent nanodiamonds were covalently functionalized with a diazoxide derivative for intracellular drug delivery. Following uptake by HeLa cells, the FNDs provided persistent fluorescence for tracking while the embedded NV centers detected drug-induced paramagnetic free radicals through $T_1$ relaxometry. The shorter $T_1$ observed for FND-diazoxide indicated enhanced local radical generation, with the response exhibiting drug-concentration dependence. This approach allowed simultaneous drug delivery and nanoscale monitoring of the local cellular response, demonstrating the potential of FNDs as quantum-enabled theranostic probes.(**Fig. 2(b)**)[75]

Another study by Miller *et al*. used these FNDs containing $NV^-$ centers as fluorescent labels in a lateral-flow assay where MW-induced modulation of $NV^-$ fluorescence combined with lock-in detection enabled background-free, ultrasensitive detection of biotin-avidin and single-copy HIV-1 RNA.[16] More recently, the work by Myzk *et al*. utilized these FND sensors to probe nanoscale redox activity within endo-lysosomal compartments of living cardiac fibroblasts, where $T_1$ relaxometry detected changes in free radicals and paramagnetic iron during fibroblast-to-myofibroblast transdifferentiation under varying substrate stiffness, cell passage number, and TGF-β stimulation.[76]

However, FNDs generally exhibit inferior spin properties compared with bulk diamond. In a nanodiamond, a large fraction of the atoms reside on or near the surface which introduces surface defects, dangling bonds, and adsorbed paramagnetic species that shorten spin coherence times and promote charge-state instability through conversion of the useful $NV^-$ state into the neutral $NV^0$ state. Consequently, maintaining stable fluorescence and long coherence times remains one of the central challenges limiting the sensitivity of FND based quantum sensors. Considerable research therefore focuses on optimizing nanodiamond synthesis, purification, and surface engineering to bridge the performance gap between bulk and nanoscale diamond platforms.

## 3.3 Heterogeneous and Hybrid Diamond Quantum Sensing Platforms

Although single-crystal bulk diamonds and these FNDs possess outstanding intrinsic quantum properties, their direct implementation in practical quantum technologies is limited by several engineering challenges. Bulk diamond is difficult to microfabricate, lacks intrinsic electrical conductivity for integrated circuitry, and suffers from significant photon losses due to its high refractive index ($n \approx 2.4$), while FNDs exhibit reduced spin coherence because of surface-induced decoherence. To overcome these limitations, hybrid diamond platforms integrate quantum-grade diamond with complementary technologies such as semiconductor electronics, photonic devices, optical fibers, and microelectromechanical systems (MEMS), thereby preserving the exceptional spin properties of NV centers while enabling scalable and multifunctional quantum devices.[7778]

Among these approaches, Diamond-on-Insulator (DoI) platforms integrated with Silicon-on-Insulator (SOI), CMOS, and MEMS technologies have attracted considerable attention.[79] These architectures enable lithographic integration of coplanar waveguides, microcoils, photodetectors, and control electronics directly adjacent to shallow NV centers.[80] The close proximity of the MW structures significantly enhances the local MW field, facilitating rapid coherent spin manipulation with reduced power consumption and minimal sample heating, which is particularly advantageous for biological applications. Furthermore, monolithic integration with semiconductor electronics facilitates on-chip signal processing, data acquisition, and compact device fabrication.[54,81]

Hybrid architectures also extend quantum sensing beyond conventional microscope-based configurations. Integration of NV-containing microdiamonds with optical fibers creates flexible endoscopic probes capable of delivering excitation light and collecting fluorescence from otherwise inaccessible environments, including deep biological tissues, blood vessels, microfluidic systems, and harsh industrial settings.[82] Similarly, incorporating shallow NV centers directly into diamond anvil cells enables in situ magnetometry, thermometry, and stress mapping under ultrahigh pressures without requiring electrical connections inside the pressure chamber.

Kim *et al*. fabricated the essential ODMR components - an on-chip microwave inductor, an optical pump-beam filter, and a photodetector - as stacked layers within a standard 65-nm CMOS process, all confined to a 200 μm × 200 μm footprint. A diamond slab containing an NV ensemble (~0.01 ppm density) is placed directly on top of the chip, with a 45°-cut edge enabling side-coupled green laser excitation that minimizes pump-laser background reaching the underlying photodetector. This CMOS-diamond hybrid platform exemplifies a distinct integration strategy compared to photonic-waveguide or pick-and-place approaches: rather than coupling NV emission into an external photonic circuit, it co-locates the entire sensing and control electronics beneath the diamond itself, directly leveraging mature semiconductor foundry processes. (**Fig. 2(c)**)[83] Du *et al*. developed a neuromorphic vision sensor to encode fluorescence changes as spikes for ODMR sensing, with such integration they have achieved a 13-fold improvement in temporal resolution while maintaining comparable precision in ODMR resonance-frequency detection relative to state-of-the-art frame-based approaches.[84]

In addition, hybrid integration with photonic waveguides, optical cavities, and plasmonic nanostructures enhances fluorescence collection efficiency and spin readout fidelity, improving the sensitivity and speed of quantum measurements.[85] By combining the superior quantum properties of diamond with the maturity of semiconductor, photonic, and fiber-optic technologies, hybrid diamond platforms provide a practical route toward compact, scalable, and high-performance quantum sensors for applications ranging from biomedical diagnostics and neuroscience to materials characterization and industrial inspection.[86]

# 4 Applications of NV magnetometry for Biosensing

NV-center based magnetometry has emerged as a versatile quantum-sensing platform for probing weak magnetic signals associated with a wide range of biological processes. Its nanoscale sensitivity, room-temperature operation, and compatibility with physiological environments have facilitated a broad range of applications, including neuronal action-potential detection, ROS sensing, nanoscale NMR, drug-delivery monitoring, wide-field cellular magnetic imaging, and biomarker detection etc. These applications exploit either resonance shifts in ODMR or changes in NV spin relaxation induced by local magnetic fluctuations. Some specific cases have been discussed in the subsequent section.

## 4.1 Neural Activity Monitoring

The NV center-based quantum magnetometry has opened up a whole new dimension for non-evasive neural activity monitoring, offering magnetic field sensitivity with high spatial resolution under ambient conditions.[87,88] Conventional electrophysiological techniques, such as patch-clamp recordings and microelectrode arrays, require direct electrical contact with neurons and are often invasive, limiting long-term measurements and large-scale neural mapping.[89,90] Optical techniques based on calcium or voltage-sensitive fluorescent indicators provide excellent spatial resolution but rely on exogenous labels and are susceptible to photobleaching, phototoxicity, and indirect reporting of neuronal activity.[91] In contrast, NV-center quantum sensors detect the weak magnetic fields generated directly by ionic currents associated with neuronal action potentials, allowing label-free and minimally invasive monitoring of neuronal dynamics. Primarily this has been realised with ODMR, where magnetic fields generated during neuronal depolarization induce Zeeman shifts in the spin resonance frequencies of shallow NV centers. By tracking continuously these shifts one can quantitatively reconstruct the temporal evolution and spatial distribution of the magnetic fields. Hall *et al*. first established the theoretical feasibility of using NV-center diamond magnetometry to image neuronal action potentials by simulating the magnetic fields generated by hippocampal neurons. Their work laid the foundation for subsequent experimental demonstrations of quantum diamond neuroimaging.[92] Hansen *et al*. investigated the neuronal action potential using a real living tissue through NV magnetometry. It is the first

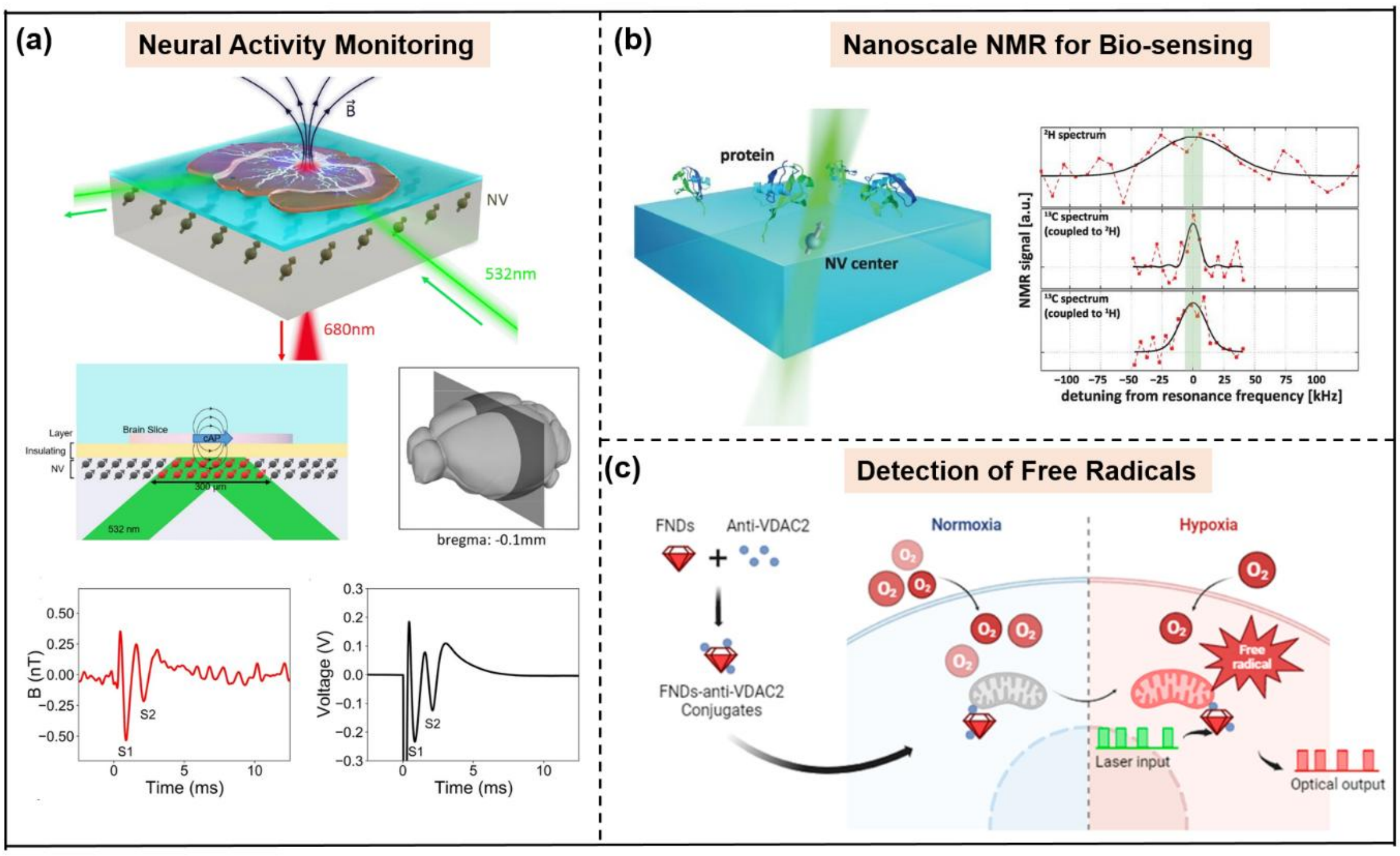

**Fig. 3. Emerging applications of NV-center-based quantum biosensing. (a)** *Neuronal activity monitoring:* NV centers in diamond facilitate non-invasive detection of weak magnetic fields generated by neuronal action potentials, facilitating label-free electrophysiological measurements and spatiotemporal mapping of neural activity. Wide-field ODMR imaging allows reconstruction of magnetic field dynamics associated with neuronal firing and brain-slice activity. *Adapted from Ref. [88]* **(b)** *Nanoscale NMR:* Schematic of single-protein NMR detection using an NV center, with ubiquitin proteins attached to the diamond surface via EDC/NHS carbodiimide cross-linking. Representative $^{2}$H and $^{13}$C NMR spectra of isotopically labeled ubiquitin, showing narrower $^{13}$C linewidths than broader, quadrupole-shifted $^{2}$H spectra, leading to single-protein detection and chemical composition analysis. *Adapted from Ref. [99]* **(c)** *Detection of ROS through $T_1$ relaxomtery:* Schematic illustrating mitochondria-targeted free-radical sensing using FNDs conjugated with anti-VDAC2 antibodies, which localize the FNDs to the mitochondrial outer membrane in H9c2 cardiomyocytes. *Adapted from Ref. [108]*

demonstration in living mammalian brain tissue. In this setup, a living brain slice or neural tissue is positioned directly over a near-surface NV ensemble in a bulk single-crystal diamond substrate.[88] Upon green optical excitation (532 nm), action potentials propagating through the axon produce microscopic ionic currents, inducing tiny magnetic field fluctuations (on the order of sub-nanoteslas) that shift the NV spin resonance frequencies, which were detected through ODMR. In this wide-field configuration, neurons or brain slices are cultured directly on the diamond surface containing a dense ensemble of shallow NV centers (5-20 nm below the surface). The NV ensemble layer were excited from below so that it does not accidentally damage the real tissue. To demonstrate its capabilities, the neurons were electrically stimulated using a bipolar electrode, while the response was measured simultaneously by a NV-diamond magnetometer (magnetic signal) as well as with Conventional Ag/AgCl electrode (electrical signal). The remarkable observation is that the magnetic signal measured by the NV centers exhibits exactly the same temporal structure as the electrical recording. This demonstrates that the NV sensor is detecting the magnetic fields generated by the same neuronal action potentials measured electrically. (**Fig. 3(a)**) Another study by Fathima *et al*. demonstrated the feasibility of FND-based quantum magnetometry in live mouse neuronal cells by achieving stable ODMR measurements and detecting externally applied magnetic fields.[93]

Continued improvements in quantum coherence, intracellular nanodiamond delivery, and hybrid sensing platforms are expected to extend NV-center magnetometry beyond proof-of-concept demonstrations toward clinically relevant applications, including real-time neuronal action potential monitoring, brain-machine interfaces, and early diagnosis of neurological diseases.

## 4.2 Nanoscale MRI

Unlike conventional MRI, which relies on the collective response of billions of nuclear spins and provides millimeter-scale spatial resolution, NV-based nanoMRI detects the magnetic fields generated by a small ensemble or even single nuclear spins located within a few nanometres of a shallow NV center.[94,95] With NV center based diamond magnetometry, the NV center itself acts as an atomic-scale quantum sensor whose spin coherence is manipulated using dynamical decoupling pulse sequences to selectively detect oscillating magnetic fields produced by nearby nuclei.[96] For bio-analytical applications, target analytes are immobilized onto functionalized diamond surfaces embedded with shallow NV centers (typically 2-5 nm beneath the surface). This near-surface proximity overcomes signal dissipation constraints, facilitating the detection of subtle nuclear magnetic signatures that lie far beyond the sensitivity limits of inductive bulk NMR.[97]

Rugar *et al*. demonstrated magnetic resonance imaging with single-NV-based proton with a spatial resolution of approximately 12 nm, establishing NV magnetometry as a powerful platform for nanoscale MRI and high-resolution molecular imaging.[98] Lovchinsky *et al.* demonstrated nanoscale NMR spectroscopy of individual ubiquitin proteins immobilized on a functionalized diamond surface containing NV-centers These shallow NV centers provided nanoscale coupling to the protein's nuclear spins, while quantum-logic-enhanced readout enabled detection of individual protons and spectroscopic identification of 1H and 13C signals. Surface functionalization played an important role here. It localized the proteins within a few nanometers of the NV layer, which maximized the magnetic coupling between the nuclear spins and the quantum sensor. This work demonstrated that NV-based nano-NMR can provide molecular-level structural and chemical information from extremely small numbers of biomolecules, overcoming the ensemble-averaging limitations of conventional NMR.[99] (**Fig. 3(b)**) More recently, Abendroth *et al*. fabricated a quantum sensor with nanopillar waveguides embedded with single NV centers, which were then were employed for $^{19}F$ NMR detection of covalently bound fluorinated molecules, achieving sensitivity at the level of approximately 100 molecules.[100]

Growing advances in NV magnetometry-based nanoscale NMR are expected to facilitate faster, higher-sensitivity detection and chemical characterization of individual biomolecules, with potential for single-protein structural analysis and real-time molecular diagnostics at nanometre spatial resolution.

## 4.3 Detection of Reactive Oxygen Species (ROS)

Reactive oxygen species (ROS), including superoxide ($O_2{\bullet}^-$), hydroxyl radicals (•OH), hydrogen peroxide ($H_2O_2$), and singlet oxygen ($^1O_2$), play crucial roles in cellular signaling, immune responses, and metabolism. However, excessive ROS accumulation is implicated in cancer, neurodegenerative disorders, and cardiovascular diseases. Therefore, the ability to monitor ROS with high spatial and temporal resolution is essential for understanding disease progression and developing effective therapeutic strategies.[101]

The conventional ROS detection techniques involves fluorescence probes, that often suffers photobleaching and limited temporal resolution.[102] NV centers detect ROS indirectly by sensing the fluctuating magnetic fields generated by the unpaired electron spins of free radicals. These stochastic magnetic fluctuations accelerate the longitudinal spin relaxation ($T_1$) of nearby NV centers. By measuring changes in the $T_1$ relaxation time, the local concentration and dynamics of ROS can be quantified in real time without perturbing the target molecules.[103,104]

FNDs based diamond platforms are very well suited for intracellular ROS sensing because of their excellent biocompatibility, chemical inertness, and photostability.[105] In particular, functionalized FNDs can be readily internalized by living cells via endocytosis or targeted to specific organelles such as mitochondria, lysosomes, or the nucleus using appropriate surface ligands.[106] Since mitochondria are the primary source of intracellular ROS, mitochondria-targeted nanodiamonds enable localized monitoring of oxidative stress with nanometre-scale spatial resolution.

Wu *et al*. demonstrates that FND-based NV magnetometry can act as a chemically inert intracellular quantum sensor for monitoring ROS-associated magnetic fluctuations in real time, facilitating non-invasive detection of radical stress in individual living cells following stimulation with SST, TPP, and TAT ligands.[107] Similarly, Fan *et al*. developed a mitochondria-targeted quantum sensing platform using anti-VDAC2-functionalized FNDs containing NV centers to monitor intracellular free-radical generation in living cardiomyocytes during hypoxia and reoxygenation. With $T_1$ relaxometry, they quantitatively detected changes in mitochondrial ROS through magnetic-noise-induced spin relaxation. The study revealed a significant burst of free-radical production upon reoxygenation, demonstrating the capability of NV-based FND sensors for real-time, label-free, and non-invasive monitoring of oxidative stress at the subcellular level.(**Fig. 3(c)**)[108]

## 4.4 Drug Delivery and Therapeutic Monitoring

NV center-bearing FNDs based platforms have emerged as powerful theranostic nanoplatforms that combine targeted drug delivery with real-time quantum sensing at the single-cell level.[75,109] These FNDs are highly bio-compatible, chemically stable and easy functionalizable surface, which can retain their spin properties inside a cell.[110] Hence, unlike other conventional drug carriers these NV centers can continuously monitor the local biochemical environment through ODMR or $T_1$ relaxometry without interfering with cellular function.[43,111]

This dual functionality aids in theranostic applications, where drug delivery and treatment response can be evaluated simultaneously. Hence, targeted FNDs can track intracellular trafficking, drug release, and cellular uptake in real time, providing valuable information on therapeutic efficacy and pharmacokinetics. The study by Wu *et al*. have utilized these FNDs for delivering the cancer drug Diazoxide (DZX) to TNBC (Triple negative breast cancer) cells. The authors functionalized FNDs with hyperbranched polyglycerol (HPG) and an acid-labile DMA linker attached to the anti-cancer drug DZX. In the mildly acidic environment of endo/lysosomes (pH~6.5), the pH-sensitive amide linkage cleaves, releasing ~ 78% of DZX within 8 hours, compared to only 26 % at physiological pH 7.4. Released DZX opens mitochondrial ATP-sensitive potassium channels, causing a steady increase in mitochondrial reactive oxygen species (ROS\$O_2^-$) over time. Using NV $T_1$ relaxometry, they discovered that

despite elevated mitochondrial ROS, local radical loads in the surrounding cytoplasm and lysosomes significantly decreased after 24 hours of treatment. Thus, these FNDs based sensing platforms allows for real-time, non-invasive assessment of drug uptake and therapeutic response. (**Fig. 4(a)**)[112] Similarly, Tian *et al*. demonstrated that FNDs containing NV centers functionalized with a diazoxide derivative, facilitates efficient cellular uptake and intracellular delivery of the anticancer drug to HeLa cells.[113]

Future developments of FND-based drug-delivery systems could facilitate targeted and controlled intracellular drug release while simultaneously exploiting NV-center sensing to monitor therapeutic responses in real time. Such multifunctional platforms may support image-guided, personalized therapy by combining localized drug delivery with in situ monitoring of cellular responses.

## 4.5 Wide-Field Cellular Magnetic Imaging

Wide-field magnetic imaging using NV centers in diamond offers a non-invasive, high-throughput platform for mapping magnetic fields and free-radical dynamics across entire cell populations.[114,115] Unlike single-point confocal scanning setups, wide-field systems project an optical field across a large field of view onto a camera (such as a CMOS or sCMOS detector), enabling parallel readout from thousands of NV centers simultaneously.[116]

Using dense ensembles of shallow NV centers in diamond substrates or internalized FNDs, wide-field imaging achieves optical diffraction-limited spatial resolution. This allows real-time visualization of chemical and magnetic heterogeneity across distinct subcellular compartments, such as lysosomes, cytoplasm, and mitochondria. Because quantum measurements can be conducted under low optical excitation powers and without thermal damage, wide-field NV magnetometry is exceptionally well-suited for living-cell studies. It enables long-term, quantitative monitoring of cellular processes-such as drug uptake, redox homeostasis, and iron storage dynamics-within living cells in real time.[117,118,119] Sage *et al*. demonstrated for the first time wide-field magnetic imaging of living magnetotactic bacteria under ambient conditions, achieving subcellular spatial resolution of approximately 400 nm using a shallow ensemble of NV centers implanted near the surface of a diamond chip.[120]

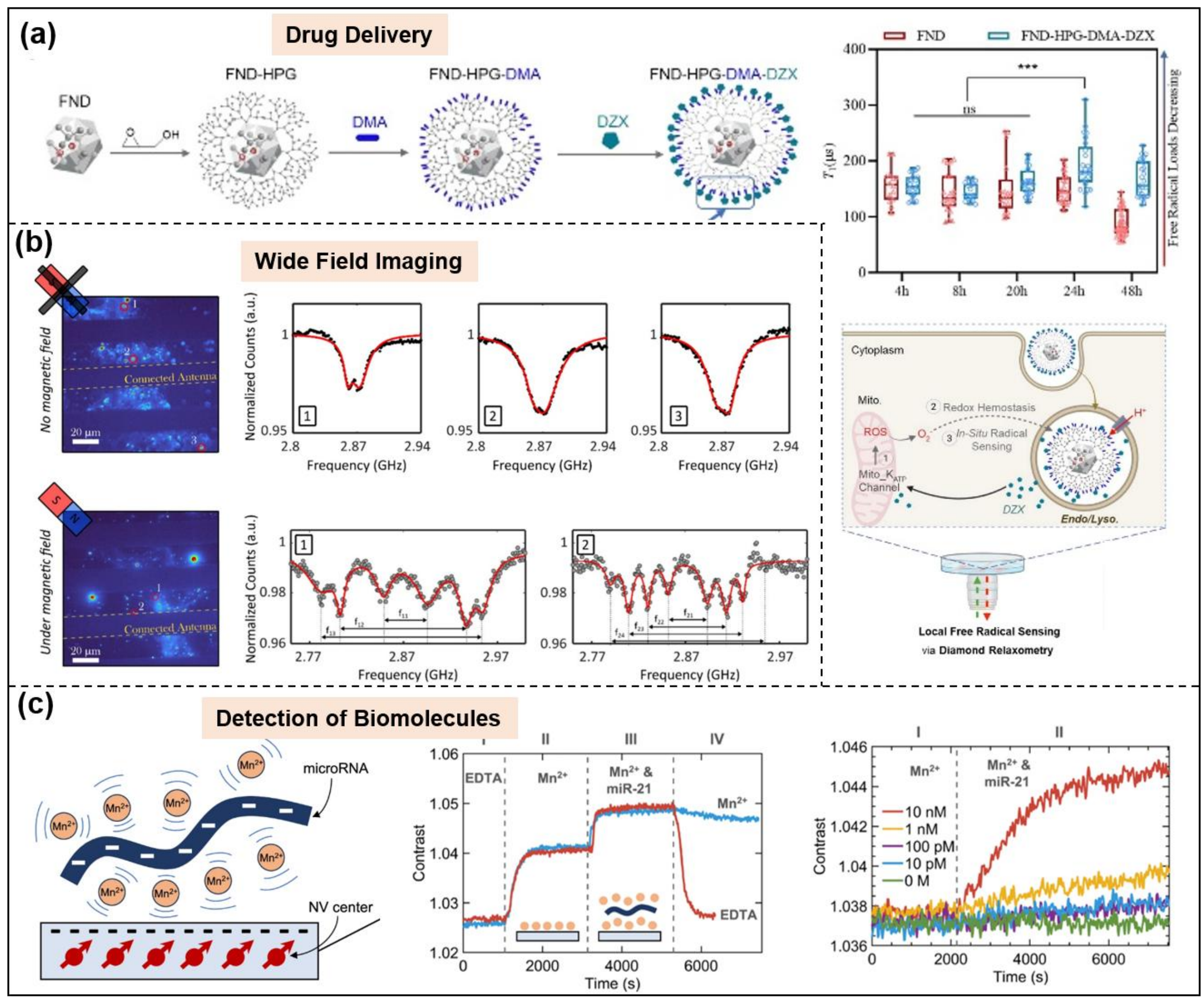

**Fig. 4. Emerging applications of NV-center-based quantum biosensing (a)** *Drug Delivery*: Stepwise functionalization of FNDs with HPG, DMA, and DZX to create a targeted theranostic probe. Following endocytosis, acidic endo-/lysosomal conditions trigger DMA cleavage and DZX release, activating mitochondrial ATP-sensitive potassium channels and enhancing ROS production. Concurrently, embedded NV centers facilitate in situ free-radical sensing via relaxometry-based detection of local magnetic noise. *Adapted from Ref. [112]* **(b)** *Wide-field ODMR imaging of biological samples*: Fluorescence images and corresponding ODMR spectra acquired in the absence and presence of an external magnetic field demonstrate spatially resolved magnetic sensing using FNDs distributed within cellular environments. Magnetic-field-induced Zeeman splitting enables extraction of local magnetic information from individual sensing regions. *Adapted from Ref. [121]* **(c)** Biomolecular sensing: Label-free detection of microRNA-21 through magnetic-noise sensing of $Mn^{2+}$ counterions accumulated near adsorbed nucleic acids. Shallow NV centers monitor changes in spin relaxation contrast arising from local magnetic fluctuations, leading to real-time and concentration-dependent quantification of microRNA under physiological conditions. *Adapted from Ref. [124]*

To demonstrate wide-field cellular magnetic imaging, Costa *et al*. performed continuous-wave

ODMR measurements on functionalized fluorescent nanodiamonds (f-FNDs) attached to SH-SY5Y neuronal cells cultured directly on a microwave antenna under total internal reflection fluorescence (TIRF) microscopy. The strong fluorescence of individual f-FNDs enabled simultaneous localization of multiple sensing sites, while ODMR spectra acquired from different regions exhibited characteristic resonance shifts and Zeeman splitting in the presence of an external magnetic field. By analysing the frequency splitting of the four possible NV crystallographic orientations, the local magnetic field at each sensing location was quantitatively determined, demonstrating the feasibility of wide-field magnetic imaging in living cellular environments using distributed nanodiamond quantum sensors.(**Fig. 4(b)**)[121]

Future developments in wide-field NV magnetometry could transform cellular biosensing by enabling high-throughput, non-invasive mapping of cellular magnetic signals and their temporal evolution, providing deeper insights into electrophysiological activity and intracellular biochemical processes.

## 4.6 Molecular Biomarker Sensing

Apart from the above application areas, the NV-center based magnetometry is widely used for label-free molecular biomarker sensing, facilitating highly sensitive detection of nucleic acids, proteins, and other disease-related biomolecules under physiological conditions. Target biomolecules are immobilized on a functionalized diamond substrate through specific recognition elements such as antibodies, aptamers, or complementary DNA probes, allowing selective and quantitative detection without the need for fluorescent labels.[122,123]

Zalieckas *et al*. utilized the $T_1$ relaxometry for the detection of microRNA-21 (miRNA-21). As shown in **Fig. 4(c)**, negatively charged microRNA-21 (miR-21) recruits $Mn^{2+}$ counterions near the diamond surface, where their fluctuating magnetic fields shorten the $T_1$ relaxation time of shallow NV centers, producing a measurable increase in relaxation contrast. The sensor exhibits a concentration-dependent response over the picomolar-to-nanomolar range, thus allowing real-time, highly sensitive quantification of microRNA biomarkers under physiological conditions without the need for fluorescent or radioactive labels.[124] Vindolet *et al*. have demonstrated that NV-center diamond exhibit enhanced sensitivity and wider detection range for detecting the VHS/G57 biomarker compared to the conventional lateral flow assays (LFA) labels.[125]

With the current progress in functionalized diamond sensors, improved sensitivity, and multiplexed wide-field readout could facilitate rapid, quantitative, and highly sensitive detection of multiple biomarkers in complex biological samples, supporting point-of-care diagnostics and real-time disease monitoring.

Beyond these key application areas, NV-center-based quantum diamond sensors are also being explored for a range of emerging biosensing applications, including intracellular temperature mapping, pH sensing, ion and metabolite detection, membrane dynamics, and monitoring of cellular metabolism etc. In **Table 1**, we summarize recent studies on biosensing using NV-center magnetometry, highlighting the different diamond platforms, sensing modalities, and target analytes investigated in each application.

**Table 1:** Summary of different NV-center diamond magnetometry platforms and their applications in biosensing.

| Application | Sensing Modality | Biological Target | Diamond Platform | Sensitivity | Contribution | Ref |
|---|---|---|---|---|---|---|
| Neuronal Action Potential Detection | ODMR | Neuronal Currents | Bulk Diamond | 50 pT/$\sqrt{Hz}$ | Recording neuronal activity in mouse brain slices using NV-diamond magnetometry for the first time | [88] |
| | ODMR | Neuronal Currents | FNDs | 1.7 mT | FND-based quantum sensing approach for recording action potentials from individual neuron cells | [93] |
| | ODMR | Neuronal Currents | Bulk Diamond | 15 pT/$\sqrt{Hz}$ | First experimental detection of single-neuron action potentials with NV centers | [87] |
| | ODMR | Neuronal Currents | Bulk Diamond | 55 μV/$\sqrt{Hz}$ | Development of diamond quantum microscope capable of imaging membrane voltage changes associated with neuronal action potentials | [126] |
| ROS Detection | $T_1$ Relaxometry | Mitochondria | f-FNDs | - | Real time monitoring of mitochondrial free radical production in living cardiomyocytes during hypoxia and reoxygenation | [108] |
| | ODMR & $T_1$ relaxometry | Intracellular free radicals | Nanogel-coated FNDs | - | Real-Time Monitoring of Free Radicals in a Single Living Cell | [107] |
| | $T_1$ Relaxometry | Free radicals | FNDs | - | Real time tracking of free radical generation during a reaction | [127] |
| Nanoscale MRI | $T_1$ Relaxometry | Single ubiquitin proteins | Electronic-grade single-crystal diamond | Single Proton Spin | First single-protein magnetic resonance spectroscopy using an NV center | [99] |
| | $T_1$ Relaxometry | Microfluidic biological samples | Shallow NV ensemble in single-crystal CVD diamond | 10-30 nT/$\sqrt{Hz}$ | Demonstration of camera-based wide-field NV-NMR microscopy | [128] |
| Drug Delivery and Therapeutic Monitoring | $T_1$ Relaxometry | Triple-negative breast cancer (TNBC) cells | FNDs | -- | pH-sensitive cancer drug delivery with real-time therapeutic monitoring | [112] |
| | $T_1$ Relaxometry | HeLa cells | f-FNDs | -- | Cellular ROS/free-radical generation induced by diazoxide treatment, monitored via NV-based T1 relaxometry to assess drug-response dynamics. | [113] |
| Wide-Field Cellular Magnetic Imaging | ODMR | Neuronal cells (CaV2.2) | FNDs | 29.8 μT/$\sqrt{Hz}$ | Demonstrated spatial mapping of magnetic fields from FNDs distributed within cellular environments. | [121] |
| | ODMR | Magnetospirillum magneticum AMB-1 | single-crystal CVD diamond | -- | First demonstration of wide-field NV magnetometry for living biological cells | [129] |
| | ODMR | Mammalian tissue | electronic-grade single crystal | 50 pT/$\sqrt{Hz}$ | First NV-diamond magnetometry of mammalian action potentials | [130] |
| Biomarker Sensing | $T_1$ Relaxometry | microRNA-21 | Electronic-grade single-crystal CVD diamond | 10 pM | label-free microRNA detection using NV-center quantum sensing | [124] |
| | $T_1$ Relaxometry | Ferritin | Single shallow NV center in a diamond nanopillar | -- | Mapped intracellular ferritin by detecting magnetic noise from ferritin iron cores | [131] |

# 5 Current Challenges and Bottlenecks in NV-Center-Based Quantum Biosensing

While NV center magnetometry offers a revolutionary paradigm for label-free nanoscale biosensing, translating this quantum platform from pristine physics laboratories to robust biological and clinical environments involves surmounting several critical physical, material, and biological bottlenecks. This section details the primary challenges restricting the current throughput, sensitivity, and widespread adoption of NV-based biosensing platforms.

## 5.1 Shallow NV Decoherence

For ultra-sensitive biosensing applications, NV centers are required to be located near the diamond surface to ensure strong coupling with target biomolecules and maximize detection sensitivity. However, reducing the NV center depth increases its exposure to surface-related noise and charge instabilities. The associated noise originates from a variety of surface defects, including dangling-bond spins, paramagnetic impurities, and charge traps that give rise to fluctuating electric fields. Collectively, these surface defects generate magnetic and electric noise that strongly couple to shallow NV centers.[132] The resulting magnetic noise accelerates spin dephasing, significantly shortening the spin coherence time ($T_2$) and limiting the quantum coherence of the defect center. Since the minimum detectable magnetic field scales inversely with the square root of the coherence time, the deterioration of $T_2$ imposes a fundamental limit on the achievable magnetic sensitivity of near-surface NV-center sensors.[133] Surface defects can also originate from the NV-center fabrication process itself. For example, NV centers created through ion implantation are often accompanied by a high density of residual lattice vacancies and other implantation-induced defects, which contribute to local magnetic and electric field noise. To mitigate these detrimental effects, a variety of surface engineering and atomic termination strategies have been developed.[134] Among them, oxygen-based surface terminations have proven particularly effective in suppressing surface-induced noise and improving the stability of shallow NV center. (**Fig. 5(a)**)

Oxygen termination is typically achieved through high-temperature annealing in an oxygen-rich atmosphere or by exposure to oxygen plasma. These treatments remove disordered $sp^2$-bonded carbon and surface contaminants, replacing them with stable oxygen-containing functional groups such as carbonyl (C=O), hydroxyl (C–OH), and ether (C–O–C) species. The resulting oxygen-terminated surface modifies the electronic structure of diamond by shifting the surface Fermi level toward lower energies and increasing the electron affinity.[135]

Consequently, the negatively charged $NV^-$ state is energetically stabilized, suppressing charge-state conversion to the neutral $NV^0$ state. This enhanced charge-state stability, together with the reduction of surface-related noise, leads to improved photoluminescence contrast, longer spin coherence times, and enhanced sensitivity in NV-based quantum sensing applications. Other than O termination, F termination is also an effective strategy as it creates a remarkably homogenous electrostatic surface layer that minimizes localized electric field gradients, hence drastically reducing $T_2$ spectral diffusion caused by charge fluctuations.[136] H-termination creates a hydrogen-induced surface dipole that drives the diamond surface toward negative electron affinity (NEA), promoting electron transfer from the diamond and making the near-surface $NV^-$ state less stable compared with O- or F-terminated surfaces.[137] In addition to surface termination strategies, surface-induced noise can also be minimized during the NV-center fabrication process itself. For example, employing low-energy ion implantation reduces lattice damage near the diamond surface, while multi-step high-temperature annealing promotes the recombination of implantation-induced vacancies and the removal of residual crystal defects.[138] These approaches lead to a lower density of paramagnetic defects and charge traps, thereby improving the coherence properties of shallow NV centers and facilitating high-precision quantum sensing. Recent studies have shown that two-dimensional (2D) capping layers such as hBN can effectively stabilize shallow NV centers by electrostatically screening local electric-field fluctuations. This dielectric screening suppresses charge-state instability and spin dephasing, resulting in improved photoluminescence contrast and longer spin coherence times.[139]

## 5.2 Refractive Index induced Light Trapping

Diamond possesses a high refractive index ($n_{diamond} \sim 2.42$), which presents a significant challenge for NV-center-based biosensing by limiting the efficiency of fluorescence collection and consequently reducing the signal-to-noise ratio (SNR). In most biosensing configurations, the diamond sensor is interfaced with biological media such as aqueous buffers, cell cultures, or tissue samples, which typically have a refractive index close to that of water ($n_{water} \sim 1.33$). The large refractive-index mismatch at the diamond-liquid interface results in a small critical angle for light transmission. Consequently, a substantial fraction of the NV-center fluorescence undergoes total internal reflection and remains trapped within the diamond rather than escaping toward the collection optics. As only a limited number of emitted photons reach the detector, the fluorescence signal is significantly reduced, directly compromising the sensitivity, signal-to-noise ratio, and overall performance of quantum sensing measurements.[140,141]

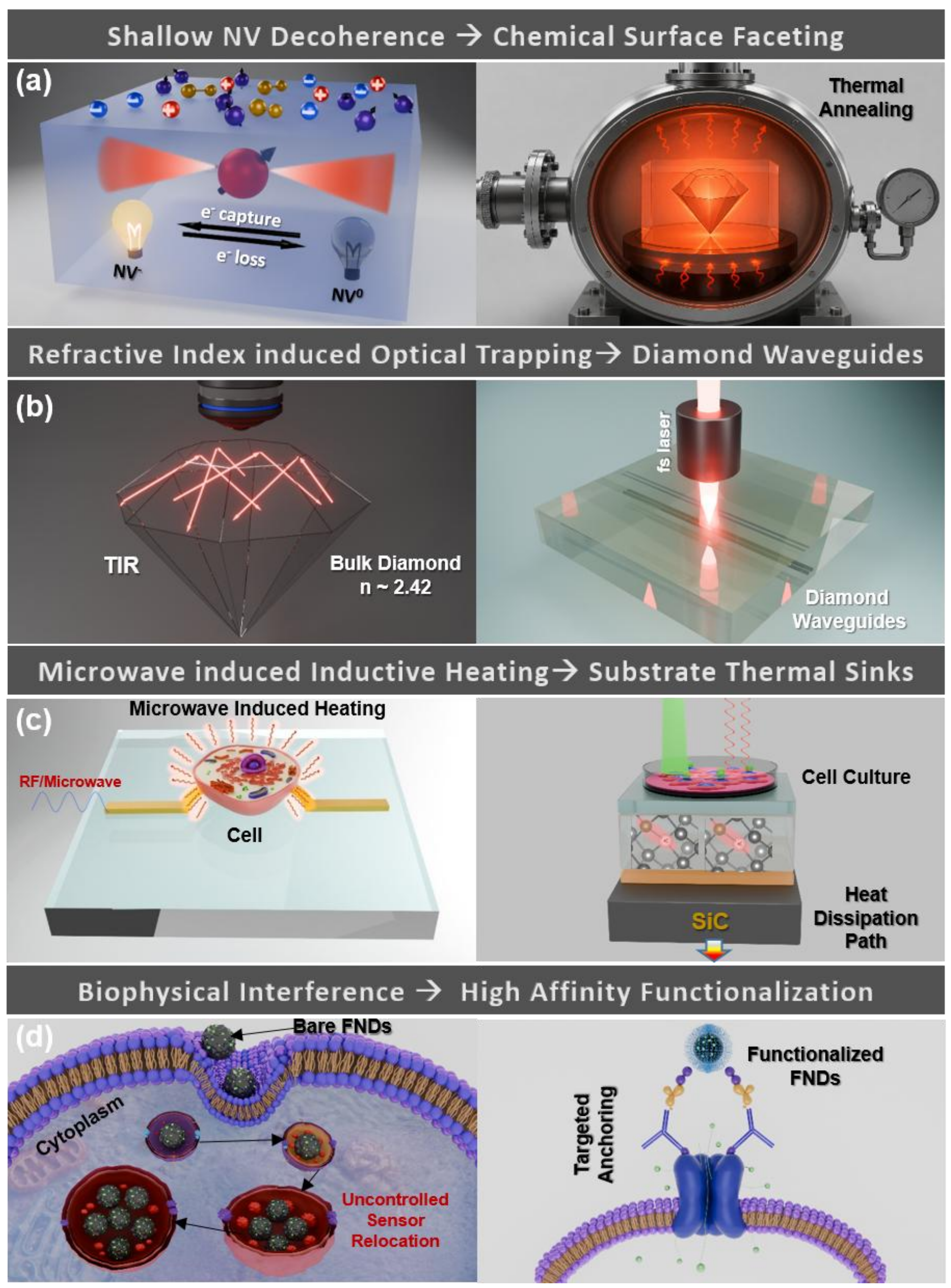

**Fig. 5. Current challenges and prospective engineering solutions in NV-diamond biosensing.** (a) Surface-induced charge instability and spin decoherence in shallow NV centers can be mitigated through surface chemical engineering and thermal annealing to stabilize the $NV^-$ state and reduce surface noise. (b) Optical losses from total internal reflection in diamond can be addressed using fs-laser-written waveguides for enhanced light extraction and routing. (c) Microwave-induced heating during ODMR can be reduced through integration of high-thermal-conductivity substrates such as SiC for efficient heat dissipation. (d) Nonspecific uptake and uncontrolled trafficking of bare FNDs can be overcome through advanced biofunctionalization for targeted localization, improved biocompatibility, and reliable intracellular sensing.

This limitation can be addressed through structural engineering of the diamond platform, particularly by employing femtosecond (fs) laser-written photonic architectures. The highly localized energy deposition of fs laser pulses enables the deterministic creation of NV centers and controlled modification of the local refractive index within the diamond. By carefully designing fs laser-written waveguides, the otherwise trapped NV fluorescence can be efficiently routed toward the collection optics, significantly improving photon extraction efficiency and signal contrast. Such integrated photonic structures offer a promising route to overcoming the optical losses associated with total internal reflection in bulk diamond.[142] Faraon *et al*. used an index-guided photonic crystal cavities in diamond to couple NV centers to high-Q cavity modes, dramatically enhancing radiative emission rates (Purcell effect) despite high index confinement. (**Fig. 5(b)**)[143]

An alternative strategy involves the use of high-numerical-aperture (NA) collection optics, which can capture a larger fraction of the emitted fluorescence and thereby enhance the signal-to-noise ratio. However, the inherently small field of view and limited working distance of high-NA objectives restrict their applicability in large-area imaging and practical biosensing configurations.[144]

## 5.3 Microwave Induced Inductive Heating

Most biological samples require an aqueous environment to remain viable. Consequently, electrolyte-rich media such as phosphate-buffered saline (PBS), Dulbecco's Modified Eagle Medium (DMEM), and physiological saline are commonly used to maintain cellular homeostasis during NV-based biosensing experiments.[145] However, a major challenge arises from the application of MW fields required to drive the spin transitions of NV centers. The oscillating electric-field component of the microwave radiation interacts strongly with the ions present in these conductive media, inducing alternating ionic currents. The resistance of the fluid to this ion motion results in Joule heating, leading to local temperature increases within the sample.[146,147]

In addition to ionic conduction losses, water molecules themselves contribute significantly to microwave absorption. The oscillating electromagnetic field continuously drives the reorientation of the polar water molecules, resulting in dielectric relaxation losses that convert electromagnetic energy directly into heat. As a result, even moderate microwave powers can produce appreciable local heating in biological samples.

Elevated temperatures can alter cellular physiology, induce stress responses, perturb biomolecular interactions, and modify the local biochemical environment. Furthermore, thermal fluctuations can shift the NV-center resonance frequencies, degrade spin coherence, and introduce measurement artifacts, thereby reducing the accuracy and reliability of quantum sensing measurements.

One promising strategy to mitigate microwave-induced heating is the use of ($T_1$) longitudinal spin-lattice relaxation sensing, which does not require continuous microwave excitation and therefore substantially reduces thermal loading of the biological sample. For applications where microwave excitation remains necessary, effective thermal management becomes essential. Recent studies have proposed the integration of high-thermal-conductivity substrates, such as silicon carbide (SiC), beneath the diamond sensor. Owing to its exceptional thermal conductivity, SiC acts as an efficient heat sink, rapidly dissipating heat generated by microwave fields and preventing its accumulation within the sensing region. (**Fig. 5(c)**)

Also, careful optimization of the microwave antenna geometry can minimize microwave-induced heating, thereby extending the measurement duration and enabling real-time monitoring of biological processes**.**

## 5.4 Biophysical Interference

This bottleneck is considered one of the most significant challenges in the development of NV-center-based biosensing platforms, particularly those employing FNDs. Owing to their nanoscale dimensions, living cells do not perceive FNDs as passive or stationary probes. Instead, they actively interact with the cellular transport machinery, leading to the internalization and sequestration of these nanoparticles within intracellular compartments. Such cellular uptake substantially compromises the sensing performance by altering the local environment of the NV centers, thereby reducing target accessibility, sensing fidelity, and the overall accuracy of the quantum readout. For applications like recording of neuronal action potentials these NV centers must reside within the evanescent electromagnetic field boundary of the target protein. But due to this endolytic trafficking the endocytosis actively pulls the FND way from the lipid bilayer, dragging it deep into the cytoplasm and since the magnetic dipole filed decays strictly as $1/r^3$, a spatial displacement > 50nm can make the quantum sensor blind to the localized event.[148] Furthermore, progressive confinement within endolysosomal compartments drives nanodiamond aggregation and compaction, bringing neighboring particles into close proximity. This enhances inter-particle electron-spin dipole–dipole interactions, which accelerate spin dephasing and shorten the coherence time ($T_2$) of the NV centers. The resulting loss of quantum coherence degrades magnetic field sensitivity and can obscure genuine environmental magnetic fluctuations, thereby compromising the accuracy and reliability of the biosensing measurement.

Several strategies have been proposed to mitigate this bottleneck, with surface engineering emerging as one of the most effective approaches. Encapsulation of FNDs with dense hyperbranched polyglycerol (HPG) or polyethylene glycol (PEG) coatings introduces substantial steric hindrance that suppresses nonspecific protein adsorption and protein-corona formation. By minimizing interactions with cellular recognition pathways, these coatings enable the nanoparticles to evade immune surveillance and significantly reduce endocytic uptake. An alternative strategy involves the conjugation of nanodiamonds with target-specific antibodies, (such as anti-$Ca_v2.2$), which selectively bind to extracellular domains of transmembrane proteins. This targeted anchoring rigidly localizes the quantum sensors at the cell membrane, preventing intracellular trafficking while preserving an optimal nanoscale separation (< 10 nm) from the biological signal source. Maintaining such close proximity is essential for maximizing magnetic-field sensitivity and ensuring high-fidelity detection of weak bioelectromagnetic signals. (**Fig. 5(d)**)

# 6 Outlook: Emerging Trends and Future Perspectives in NV-Center Quantum Magnetometry for Biosensing

The rapid advancement of NV center technology has expanded its role beyond proof-of-concept demonstrations toward practical and scalable biosensing platforms. Recent developments focus on overcoming the limitations associated with optical collection efficiency, sensor localization, molecular specificity, and real-time operation in biologically relevant environments. In this section we will discuss some of the key research areas in NV center based biosensing that are being widely developed for next generation biosensing applications.

## 6.1 Surface Functionalization

Bio-functionalization of diamond has emerged as an extremely promising direction for highly selective and target-specific diagnostics platform capable of bypassing ensemble averaging of traditional assays. Since, the weak magnetic field generated by targeted biomolecules decay very rapidly with distance (~1/r³), maximizing spatial coupling requires near-surface NV centers and direct molecular functionalization of the diamond surface. One of the major

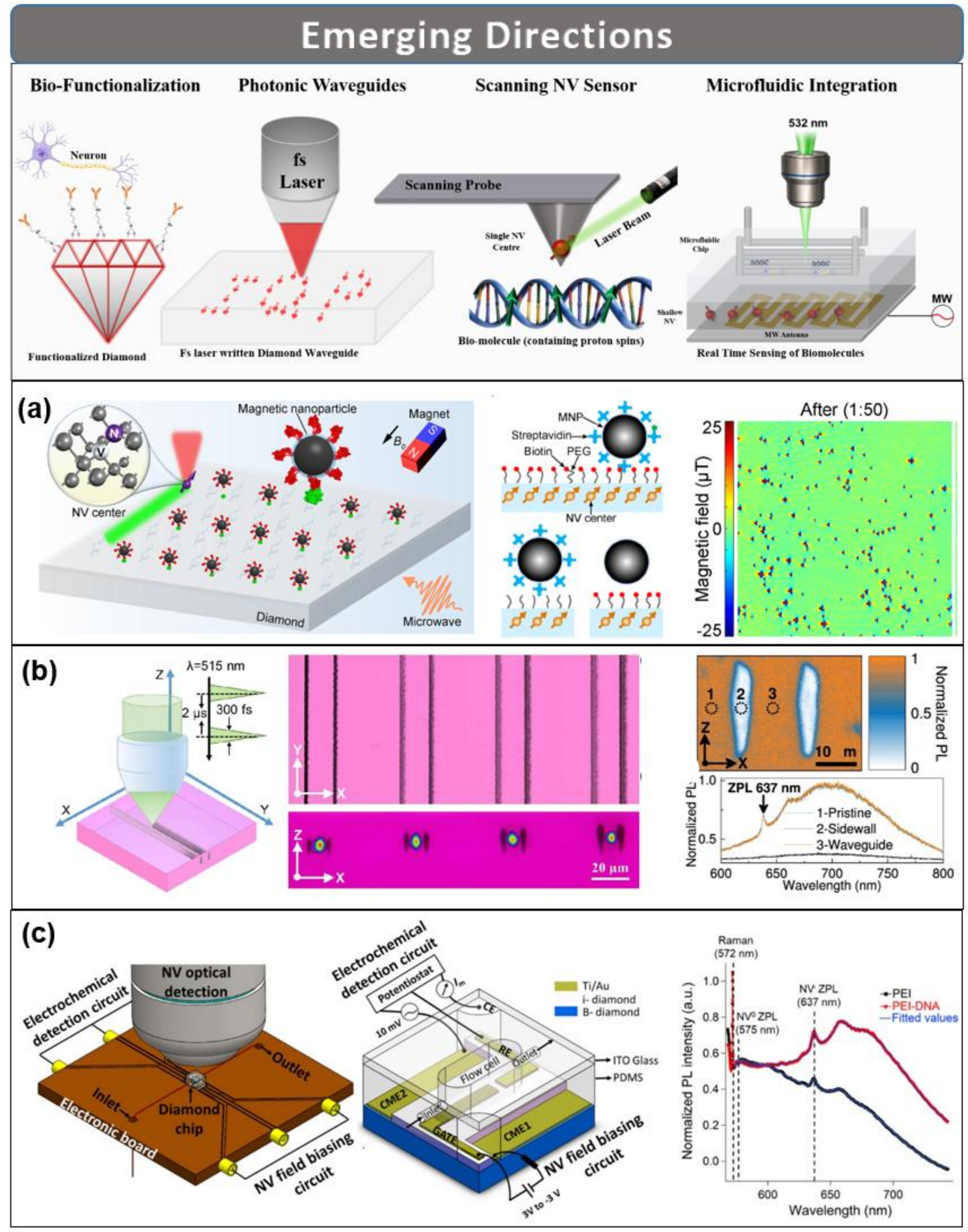

**Fig. 6. Emerging directions on NV-center based Biosensing:** *Top:* Schematic overview of next-generation diamond biosensing platforms. **(a)** Biomolecular detection using surface-functionalized MNPs and wide-field NV-center magnetometry. *Adapted from Ref. [149]* **(b)** Femtosecond-laser fabrication of diamond waveguides enabling efficient optical addressing and fluorescence collection from embedded NV centers. *Adapted from Ref. [156]* **(c)** Integrated diamond biochips combining NV quantum sensing with electrochemical and microfluidic platforms for multimodal biomolecular detection and lab-on-chip diagnostics. *Adapted from Ref. [164]*

advantages of functionalization is target specific recognition. To achieve this, the diamond surface has to be modified to present reactive handles that does not degrade the NV centers quantum coherence ($T_2$ time) or its optical readouts via ODMR. The functionalization of diamond surface can be done through various processes depending on the specific requirement of the application. For instance, oxygen based functionalization often improves the stability of the $NV^-$ charge center. This is particularly useful with FNDs where the NV center resides close to the surface and is prone to external charge fluctuations. Additionally, unfunctionalized FNDs gets aggregated very easily in high-salt physiological buffers due to charge instability, losing their tracking utility. Covering the FNDs surface with bio-polymers (such as PEG) or lipid bilayers prevents aggregation and thus increases colloidal stability and more importantly it keeps the NV sensor operating reliably inside the living cell.

Chen *et al*. developed a digital magnetic biosensing strategy based on shallow NV centers and functionalized magnetic nanoparticles (MNPs), where individual MNPs act as magnetic labels for biomolecular recognition. As illustrated in **Fig. 6(a)**, the MNPs are functionalized with biorecognition molecules such as streptavidin, allowing them to selectively bind complementary molecules immobilized on the diamond surface. Upon application of an external magnetic field, the bound MNPs generate localized magnetic stray fields, which are detected by the near-surface NV ensemble through ODMR-based magnetic imaging. The resulting magnetic-field maps enable individual nanoparticles to be spatially resolved and digitally counted, providing a direct measure of biomolecular binding events. Importantly, this approach achieved single-particle detection and digital quantification of biomolecular interactions**,** demonstrating the potential of surface functionalization in NV magnetometry for highly sensitive, label-based biosensing.[149]

In future, this strategy could be judiciously exploited to regulate the NV charge state and spin coherence during biomolecular interactions through dynamic and stimuli-responsive surface functionalization of NV-diamond. Such engineered interfaces could lead to label-free mapping of molecular binding, membrane dynamics, and intracellular magnetic processes with enhanced spatial and temporal resolution using quantum magnetometry.

## 6.2 Photonic Waveguides

The extremely high refractive index of diamond (~2.42) has made this material truly unique however, this possess a critical challenge in terms of collecting the scattered photons as 90% of the scattered photons never leave out the diamond surface due to the TIR faced at the diamond-air interface. This severely affects the magnetic field sensitivity of the ODMR and relaxometry measurements. Directly machining three-dimensional optical structures in bulk diamond circumvents this limitation. Among all other photonic strategies, fs laser writing has emerged as one of the most powerful method for integrated quantum sensing.[150] Unlike other surface fabricated optical structures, fs laser writing enables direct three-dimensional modification of the diamond interior through highly localized nonlinear absorption.[151] Ultrashort laser pulses, typically ranging from 100 to 300 fs, induce permanent refractive index modifications inside the bulk crystal without requiring lithographic processing.[152] This allows buried optical waveguides to be fabricated with micrometre precision while preserving the surrounding crystal quality and maintaining the coherence properties of nearby NV centers. The main advantage of such high energy laser writing is that it can confine the incident light into a small predefined volume and can guide the red fluorescence photons effectively into the collecting objective, hence increasing the collection efficiency, that results in increased ODMR contrast.[153] However, due to the extreme hardness and wide bandgap of diamond direct positive refractive index modification is difficult. Instead there has been reports on Type-II waveguides where two laser induced damage tracks create stress induced refractive index changes in the region between them.[154,155] Guo *et al*. developed a fs-laser-written integrated photonic diamond platform to enhance the optical excitation and collection of NV centers for quantum magnetometry. The authors fabricated buried optical waveguides inside a CVD diamond containing a high NV concentration (~4.5 ppm) by fs-laser writing, as illustrated by the waveguide cross-sectional and mode-profile images in **Fig. 6(b)**.[156] The waveguide confines and guides the excitation and NV fluorescence along the diamond, allowing a substantially larger volume of NV centers to be interrogated compared with conventional confocal excitation. Importantly, the photoluminescence spectra show that the laser-written waveguide preserves the characteristic NV emission, including the 637 nm zero-phonon line, indicating that the fabrication process does not significantly degrade the optical properties of the NV centers. This architecture therefore provides a pathway toward compact, fiber-integrated NV sensors**,** with the potential to combine the diamond chip with microfluidics and biological samples while reducing direct optical exposure. The reported device achieved 63 $pT.Hz^{-1/2}$ DC and 20 $pT.Hz^{-1/2}$ AC magnetic-field sensitivity-more than an order-of-magnitude improvement over the conventional confocal configuration.

## 6.3 Microfluidic Integration

The integration of microfluidic technology with NV-center-based quantum magnetometry has emerged as a promising strategy for developing compact, automated, and high-throughput biosensing platforms.[157] While conventional NV magnetometry is generally confined to static samples deposited directly on the diamond surface. Coupling this platform with microfluidics instead allows minute liquid volumes to be precisely manipulated and delivered to the sensing region in a controlled manner.[158] Such integration minimizes sample consumption, improves measurement reproducibility, and facilitates continuous monitoring of biochemical processes under physiologically relevant conditions.[159,160]

Typically, such microfluidic channels are fabricated with optically transparent and biocompatible polymers such as polydimethylsiloxane (PDMS) then can be placed or bonded directly over the diamond chips containing layer of shallow NV centers.[161] Since the magnetic field generated by the external spin sources decays rapidly with distance ($\propto 1/r^3$), it is ensured that that when these analyte solution flows through these channels the target molecues or cells are within a few micrometers of the NV sensing layer. Such architechture ensures efficient coupling between the biological specimen and the NV centers while simultaneously providing a controlled chemical environment. This allows real time monitoring of the dynamical biological processes including enzymatic reactions, cellular metabolism, oxidative stress, and drug responses, without repeatedly preparing fresh samples. Another significant advantage of microfluidic platforms is their compatibility with lab-on-a-chip technologies.[162] Integration of NV-diamond sensors with microfluidics, MW delivery structures, optical waveguides, and CMOS electronics has led to the development of highly compact quantum biosensing devices capable of automated sample handling and portable operation.

Lim *et al.* first that demonstrated localized NV magnetometry inside a microfluidic environment. They manipulate magnetic particles through the microfluidic device and achieve approximately 48 nm spatial precision.[163] Krečmarová *et al.* developed a label-free, microfluidic diamond biosensor that integrates shallow NV centers with electrochemical control for detecting biomolecular interactions. As illustrated in **Fig. 6(c)**, the platform combines a diamond chip containing near-surface NV centers, integrated electrical contacts for NV optical detection and field biasing, and a PDMS microfluidic flow cell with electrochemical electrodes, allowing controlled delivery of analytes directly over the sensing region. The sensing principle relies on electrical manipulation of the NV charge state: an applied bias establishes the initial NV charge-state population, while charged biomolecules introduced through the microfluidic channel modify the local surface electrostatic potential, leading to a measurable change in the NV photoluminescence. In particular, the authors demonstrated label-free DNA detection, combining the optical NV response with electrochemical measurements to validate biomolecular binding.[164] More recently, Briegel *et al*. demonstrated wide-field NV-NMR microscopy integrated with microfluidic channels, facilitating spatially resolved detection of nuclear spins from very small sample volumes. This provides a promising platform for high-throughput, non-invasive biosensing and chemical analysis, with potential for studying biological processes in microfluidic and cellular environments.[165]

Despite the tremendous promise of microfluidic-integrated NV magnetometry, challenges including precise sample positioning near the shallow NV layer, microwave-induced heating, optical losses, material biocompatibility, and scalable device fabrication must be overcome for widespread clinical translation.

## 6.4 Scanning NV Magnetometry

Beyond these techniques, scanning NV magnetometry has opened up new avenues in NV-center-based quantum sensing, leading to high-resolution spatial mapping of magnetic fields at the nanoscale. In contrast to ensemble-based wide-field NV imaging, which uses a dense layer of shallow NV centers to simultaneously probe a large field of view, scanning NV magnetometry instead relies on a single NV center located at the apex of a diamond nanopillar or an atomic force microscope (AFM) tip.[24] By raster scanning the probe across the sample surface with a tip-to-sample separation of just a few nanometers, this technique attains remarkable magnetic sensitivity along with spatial resolution that surpasses the optical diffraction limit. This allows for direct imaging of magnetic nanoparticles, magnetotactic bacteria, ferritin complexes, and the magnetic signatures of individual cells, all with nanometer-scale precision. This technique can also be applied to explore detecting single proteins, DNA molecules, and biomolecular spin labels through nanoscale NMR spectroscopy, offering the prospect of structural characterization of individual biomolecules without the need for crystallization or ensemble averaging.

Wang *et al*. demonstrated the first application of scanning NV-center magnetometry for label-free imaging of endogenous biomolecules by mapping intracellular ferritin in HepG2 cells using NV $T_1$ relaxometry. Ultrathin cell sections were scanned over shallow NV centers embedded in diamond nanopillars, where magnetic noise from the ferritin iron cores shortened the NV spin relaxation time, facilitating reconstruction of nanoscale magnetic images.[131] The magnetic maps showed excellent agreement with transmission electron microscopy (TEM), establishing scanning NV magnetometry as a powerful tool for nanoscale intracellular biosensing and magnetic imaging. In a recent study, Mosavian *et al*. demonstrated nanoscale magnetic microscopy of individual 30 nm iron-oxide nanoparticles using NV centers, achieving a spatial resolution of approximately 100 nm. This work highlights the potential of NV-based scanning magnetic microscopy for biosensing, particularly for the highly localized and spatially resolved detection of magnetic nanoparticle labels and, consequently, sensitive biomarker detection.[166]

Despite the impressive capabilities, several technical hurdles currently constrain broader biological applications. Prolonged image acquisition times, the difficulty of sustaining a stable nanometer-scale tip-to-sample separation within liquid environments, restricted imaging throughput, and the added complexity of integrating AFM instrumentation with microwave and optical excitation all remain substantial obstacles. More recently, NV center integrated onto a fiber tip instead of a conventional AFM-based scanning tip has shown tremendous potential in magnetometry based biosensing applications. This provides a compact and flexible platform for next generation medical diagnostics. The optical fiber can deliver the excitation laser and collect NV fluorescence, reducing the need for bulky free-space optics. Unlike a conventional AFM cantilever, the fiber geometry can be engineered for operation in liquid environments, making it attractive for cellular and biomolecular sensing. Kuwahata *et al*. demonstrated a compact fiber-based NV magnetometer for detecting magnetic nanoparticles, achieving nT-level magnetic sensitivity and micromolar MNP detection at millimetre-scale distances.[167] Another work by Li *et al*. also fabricated a fiber-coupled scanning NV magnetometer based on a diamond nanobeam, facilitating efficient through-fiber excitation and readout while retaining the nanoscale scanning-probe geometry.[168] In the foreseeable future scanning NV platforms are expected to enable real-time nanoscale imaging of intracellular magnetic phenomena, molecular interactions, and biochemical dynamics, establishing this technique as a key component of next-generation quantum biosensing and single-molecule biophysics.

## 6.5 On-Chip Integration of NV Magnetometry for Biomedical Applications

The translation of NV-center-based quantum magnetometry from laboratory demonstrations to practical biomedical devices requires compact, robust, and scalable sensing architectures.[169] On-chip integration brings together diamond quantum sensors with microfabricated microwave circuitry, photonic components, and microfluidic platforms to create miniaturized devices capable of real-time biosensing under physiological conditions. Such integrated platforms improve photon collection efficiency, microwave delivery, sensing stability, and device portability, thereby facilitating high-throughput measurements and point-of-care diagnostics.[170] Recent advances in microfabrication have enabled seamless integration of diamond sensors with complementary semiconductor technologies, opening new opportunities for portable quantum biosensing and clinical translation.[171,172,173]

Developing a biocompatible, chip-integrated platform for long-term quantum sensing of living biological systems requires careful consideration of several factors, including thermal management, microwave delivery, cell viability, and the choice of an appropriate diamond sensing platform. Shanahan *et al*. addressed these challenges by developing the Quantum-Biocompatible Integrated Chip (Q-BiC), which integrates a shallow NV-center diamond, a lithographically fabricated microwave antenna, and PDMS-based microfluidic chambers into a compact lab-on-chip architecture. The platform supported the long-term culture of HeLa cells and Caenorhabditis elegans (C. elegans) directly on the diamond surface while preserving cell viability and organism motility, demonstrating excellent biocompatibility. To overcome microwave-induced heating, the authors employed thermal simulations to optimize the antenna geometry, achieving efficient spin manipulation with minimal temperature rise. In addition, integrated PDMS wells and microfluidic channels enabled continuous culture under physiological conditions. This work represents an important milestone toward the realization of fully integrated, biocompatible quantum sensing platforms, capable of long-term investigations of living cells and organisms for future biomedical and lab-on-chip applications.[174] (**Fig. 7(a)**)

Webb *et al*. demonstrated an integrated on-chip NV-center magnetometry platform for non-invasive electrophysiological measurements by positioning an optogenetically stimulated mouse extensor digitorum longus (EDL) muscle directly above a CVD-grown diamond containing an NV ensemble. Continuous-wave ODMR performed through the integrated diamond chip detected transient magnetic fields (~250 pT) generated by muscle action potentials, while simultaneous electrical recordings validated the measurements. This work established the feasibility of chip-scale diamond quantum sensors for contact-free

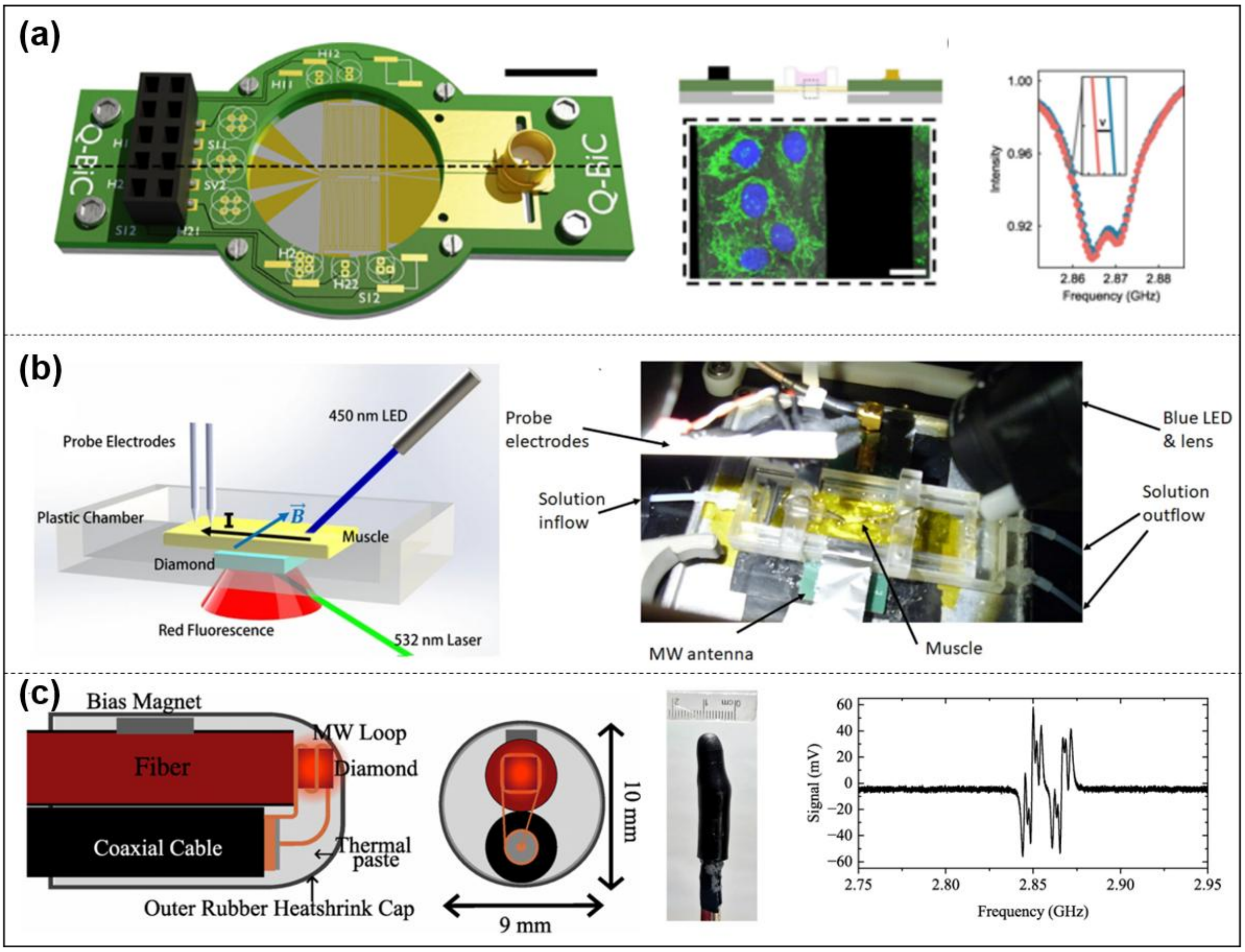

**Fig. 7. On-chip integration of NV-diamond quantum sensors for biomedical applications. (a)** Microfabricated NV-diamond biochip integrated with microwave circuitry for wide-field ODMR imaging, leading to spatially resolved magnetic sensing and cellular bioimaging on a chip-scale platform. *Adapted from Ref. [174]* **(b)** Diamond-based quantum sensing platform for electrophysiological measurements, demonstrating detection of bioelectric currents and associated magnetic fields from excitable tissues using integrated optical excitation, microwave control, and microfluidic interfaces. *Adapted from Ref. [130]* **(c)** Miniaturized fiber-coupled NV-diamond magnetometer featuring an integrated microwave loop and compact probe architecture for portable and endoscopic magnetic sensing, together with representative ODMR spectra obtained from the device. *Adapted from Ref. [175]*

bioelectromagnetic sensing, representing an important step toward integrated quantum biochips for real-time electrophysiological monitoring.[130] **(Fig. 7(b)**)

Driven by advances in miniaturization and integrated photonics, NV-center-based magnetometers are now being developed as fiber-coupled endoscopic probes for minimally invasive quantum sensing. Newman *et al*. designed a compact fiber-coupled magnetometer with a sensor head diameter of only 10 mm, consisting of a 0.5 $mm^3$ diamond containing an ensemble of NV centers, an optical fiber for laser excitation and fluorescence collection, a miniature microwave loop for ODMR, and a permanent bias magnet. This compact design enables magnetic sensing in confined anatomical regions inaccessible to conventional bulk diamond systems. The compact design is particularly attractive for image-guided surgery, sentinel lymph node mapping, endoscopic diagnostics, and future catheter-based quantum biosensing, representing an important step toward clinical translation of NV-center quantum magnetometers.[175] (**Fig. 7(c)**)

Overall, on-chip integration of NV-center magnetometry with photonic, microfluidic, and electronic components offers a promising pathway toward compact, automated, and multifunctional quantum biosensors. Continued advances in device integration, thermal management, and scalable fabrication will be crucial for translating these platforms into portable and real-time biomedical sensing technologies.

# 7 Conclusion

In conclusion, we have provided an overview of state-of-the-art NV-center-based quantum magnetometry and its growing applications in biosensing, discussing the underlying physics, various sensing modalities, diamond platforms, and their respective advantages and limitations. We have also highlighted the challenges associated with translating NV-based sensing to practical and industrial applications, including surface-induced decoherence, charge-state instability, and limited sensitivity in biological environments. Nevertheless, rapid advances in diamond engineering, surface functionalization, photonic integration, scanning-probe architectures, and microfluidic platforms are steadily improving sensor performance and expanding their applicability. This progress is also reflected in the growing industrial landscape, with companies such as QDTI, Type-I Technologies, Qnami, Diatope, QT Sense, QDSENSE, Kwan-Tek, and Bosch Quantum Sensing developing commercial technologies based on NV-center and diamond quantum sensing. These developments demonstrate that NV-center technology has moved beyond the confines of academic laboratories and is increasingly approaching industrial scale. With continued technological advances, NV-based biosensors hold significant promise for next-generation medical diagnostics, real-time cellular imaging, biomarker detection, and precision medicine, potentially transforming quantum sensing from a laboratory technology into a practical clinical and industrial tool.

# Cover Image

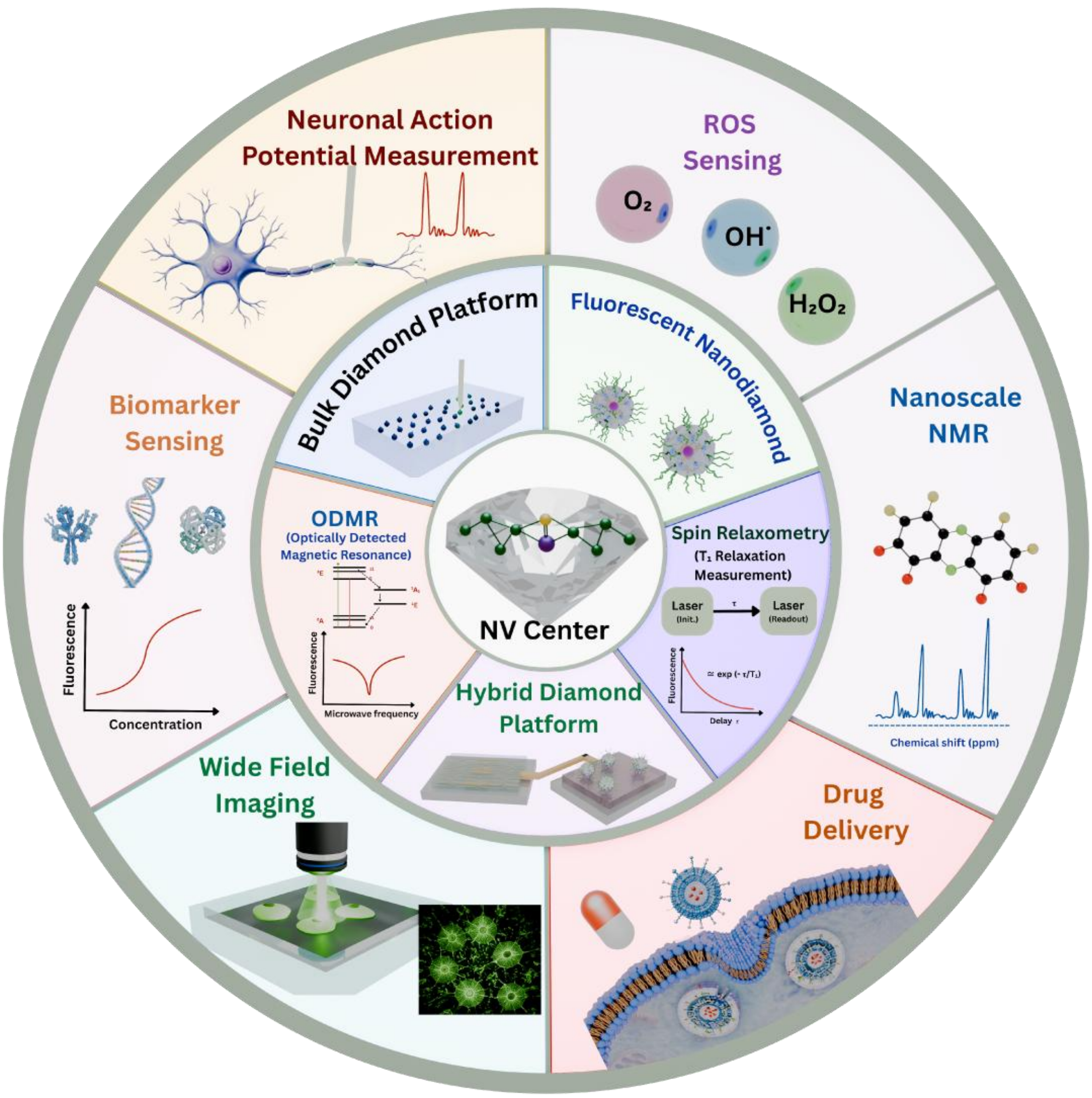

Neuronal Action Potential Measurement
ROS Sensing
$O_2$
$OH^\cdot$
$H_2O_2$
Bulk Diamond Platform
Fluorescent Nanodiamond
Nanoscale NMR
Chemical shift (ppm)
Biomarker Sensing
Fluorescence
Concentration
ODMR
(Optically Detected Magnetic Resonance)
Fluorescence
Microwave frequency
NV Center
Spin Relaxometry
($T_1$ Relaxation Measurement)
Laser (Init.)
Laser (Readout)
Fluorescence
Delay
Hybrid Diamond Platform
Wide Field Imaging
Drug Delivery